\documentclass[lettersize,journal]{IEEEtran}

\usepackage{array}
\usepackage{stfloats}
\usepackage{url}
\usepackage{verbatim}
\usepackage{amsmath,amssymb,amsfonts}
\usepackage{algorithm,algpseudocode}
\usepackage{soul,color}
\usepackage{cite}
\usepackage{textcomp}
\usepackage{titlesec}
\usepackage{indentfirst}
\usepackage{subfigure}
\usepackage{multirow}
\usepackage{hhline}
\usepackage{graphicx}
\usepackage{subcaption}
\makeatletter
\let\NAT@parse\undefined
\makeatother
\usepackage[shortlabels]{enumitem}
\usepackage[colorlinks,citecolor=green,urlcolor=blue,bookmarks=false,hypertexnames=true]{hyperref}

\begin{document}


\title{Robust Phantom-Assisted Framework for Multi-Person Localization and Vital Signs Monitoring Using MIMO FMCW Radar}
\author{Yonathan Eder, \IEEEmembership{Graduate Student Member, IEEE}, Emma Zagoury, Shlomi Savariego, Moshe Namer, Oded Cohen, and Yonina C. Eldar, \IEEEmembership{Fellow, IEEE}  \vspace{-0.7cm}
\thanks{
This research was supported by the European Research Council (ERC) under the European Union’s Horizon 2020 research and innovation program (grant No. 101000967), by the Israel Science Foundation (grant No. 536/22), and by The Manya Igel Centre for Biomedical Engineering and Signal Processing. \textit{(Corresponding author: Yonathan Eder.)}\\
The authors are with the Faculty of Mathematics and Computer Science, Weizmann Institute of Science, Rehovot, Israel (e-mail: yoni.eder@weizmann.ac.il, emma.zagury@weizmann.ac.il, shlomi.savariego@weizmann.ac.il, namer@technion.ac.il, oded.cohen@weizmann.ac.il, yonina.eldar@weizmann.ac.il).}   }

\maketitle

\begin{abstract}
With the rising prevalence of cardiovascular and respiratory disorders and an aging global population, healthcare systems face increasing pressure to adopt efficient, non-contact vital sign monitoring (NCVSM) solutions. This study introduces a robust framework for multi-person localization and vital signs monitoring, using multiple-input-multiple-output frequency-modulated continuous wave radar, addressing challenges in real-world, cluttered environments. Two key contributions are presented. First, a custom hardware phantom was developed to simulate multi-person NCVSM scenarios, utilizing recorded thoracic impedance signals to replicate realistic cardiopulmonary dynamics. The phantom's design facilitates repeatable and rapid validation of radar systems and algorithms under diverse conditions to accelerate deployment in human monitoring. Second, aided by the phantom, we designed a robust algorithm for multi-person localization utilizing joint sparsity and cardiopulmonary properties, alongside harmonics-resilient dictionary-based vital signs estimation, to mitigate interfering respiration harmonics. Additionally, an adaptive signal refinement procedure is introduced to enhance the accuracy of continuous NCVSM by leveraging the continuity of the estimates. Performance was validated and compared to existing techniques through $12$ phantom trials and $12$ human trials, including both single- and multi-person scenarios, demonstrating superior localization and NCVSM performance. For example, in multi-person human trials, our method achieved average respiration rate estimation accuracies of $94.14\%$, $98.12\%$, and $98.69\%$ within error thresholds of $2$, $3$, and $4$ breaths per minute, respectively, and heart rate accuracies of $87.10\%$, $94.12\%$, and $95.54\%$ within the same thresholds. These results highlight the potential of this framework for reliable multi-person NCVSM in healthcare and IoT applications.
\end{abstract}


\section{Introduction}
\IEEEPARstart {V}ital sign monitoring is fundamental to modern healthcare, providing essential insights into a patient’s physiological state. It plays a critical role in assessing acute conditions, enabling early detection of chronic diseases and deterioration, and improving overall patient management \cite{brekke2019value}. However, despite technological advances in medicine, traditional methods for monitoring heart rate (HR) and respiration rate (RR) remain limited. These techniques often require direct physical contact, which may cause discomfort, increase the risk of disease transmission, restrict mobility, and require full patient cooperation \cite{thakor1985electrode,ni2021automated}. With the rising prevalence of cardiovascular and respiratory disorders and the global aging population, healthcare systems face increasing strain from overcrowded hospitals and clinics. This highlights the pressing need for robust, non-contact monitoring solutions that can alleviate the burden on healthcare providers, reduce costs, and improve the quality of medical care  \cite{brownsell1999future,boric2002wireless,nangalia2010health,ceballos2015psychosocial}.

Remote sensing technologies such as millimeter-wave (mmWave) radars are particularly well-suited for non-contact vital sign monitoring (NCVSM), as they do not require users to wear, carry, or interact with any electronic device \cite{fioranelli2019radar}. These systems can detect subtle body movements associated with respiration and cardiac function. Their affordability, durability, low power consumption, and compact size make them ideal for integration into healthcare and Internet of Things (IoT) applications, facilitating seamless connectivity between patients and healthcare providers, either directly or through cloud-based infrastructure \cite{swaroop2019health}.

Initially, continuous wave (CW) radars in single-input-single-output (SISO) setups, were proposed for remotely measuring respiratory and cardiac chest movements \cite{xiao2006ka,gu2012accurate,zhao2018noncontact}. While these radars are highly sensitive and energy-efficient, their poor ability to distinguish between targets and clutter limits their practical use. In recent years, mmWave frequency-modulated continuous wave (FMCW) radars have garnered significant interest due to their ability to spatially separate objects while tracking millimeter-scale displacements \cite{alizadeh2019remote,sacco2020fmcw,antolinos2020cardiopulmonary,kim2018low,turppa2020vital}. This capability positions FMCW radars as promising candidates for multi-person NCVSM, even when employing a single channel in a SISO configuration, provided the subjects are located at distinct radial distances from the radar \cite{eder2023sparsity}.

To extend functionality for persons at the same radial distance, angular separation was incorporated by using a single-input-multiple-output (SIMO) setup, leveraging phase shifts induced by multiple receivers \cite{eder2023sparse}. Further enhancements can be realized with multiple-input-multiple-output (MIMO) configurations \cite{li2021through,feng2021multitarget,juan2022distributed,hsieh2024multiperson}, which may offer higher signal-to-noise ratios (SNR) and improved spatial resolution. The latter is achieved by employing orthogonal signals from multiple transmit antennas, using techniques such as time-division multiplexing (TDM) \cite{li2021signal,xu2022simultaneous}. In a uniform linear array (ULA), by transmitting and receiving independent signals over a common signal path, this approach creates a larger virtual antenna array, whose effective size equals the product of the number of transmit and receive antennas, ultimately enhancing the radar’s angular resolution.

To integrate radar-based multi-person NCVSM into real-world crowded environments such as waiting areas or emergency rooms, methods must be accurate and robust against the inherent challenges of these settings. These challenges stem from factors such as low signal-to-noise ratio (SNR), strong clutter reflections, multi-path effects, random body movements (RBMs), and interfering harmonics caused by the non-pure sinusoidal nature of thoracic activity \cite{mercuri2021enabling,hsieh2024multiperson,eder2023sparsity,paterniani2023radar}. Developing such robust methods requires both the design of advanced algorithms that leverage domain-specific knowledge of the data as well as a proper validation scheme capable of reliably assessing their performance. 

Traditionally, algorithms for multi-person NCVSM using SIMO or MIMO FMCW radar involve four fundamental stages: 1. Preprocessing: In this stage, the In-phase (I) and Quadrature (Q) channels of each receiver are processed to structure the data for vital signs monitoring \cite{alizadeh2019remote,sacco2020fmcw,antolinos2020cardiopulmonary,mercuri2019vital}. However, the I/Q framework is limited by imperfections such as non-orthogonality and channel gain mismatches, which distort vital sign extraction and reduce accuracy \cite{fatadin2008compensation,mahendra2020compensation}. Our previous work has demonstrated the feasibility of exploiting only a single channel to address these limitations \cite{eder2023sparsity,eder2023sparse}.

2. Human Localization: Accurate localization of individuals' thoraces is essential for extracting their vital signs. Several studies employed the \textit{angle-FFT} technique \cite{ahmad2018vital,gao2019experiments,han2021detection}. While simple and computationally efficient, it is highly susceptible to reflections from static objects and offers relatively poor angular resolution \cite{gao2019experiments}. Some researchers suggested the Multiple Signal Classification (MUSIC) algorithm \cite{konno2014experimental,kim2020low,hsieh2024multiperson} to enhance angular resolution. However, it is highly sensitive to dynamic clutter, such as oscillating objects, which may lead to potential errors in realistic scenarios. To improve the separation of objects, the authors in \cite{xiong2022vital} extended MUSIC by a second-order differential approach, whereas in \cite{wang2021multi}, an adaptive beamforming method was adopted. Nevertheless, both remain challenged by closely spaced targets in cluttered settings. A frame-averaging technique was suggested in \cite{wang2021driver} to mitigate the effect of static clutter, although it is sensitive to parameter selection and the presence of active clutter.

3. Vital Doppler Extraction: Once the chest wall of the humans is accurately located, the next step involves extracting their phase-modulated thoracic vibrations to facilitate subsequent NCVSM. This process is commonly achieved through beamforming \cite{xiong2022vital,ahmad2018vital}, followed by arctangent demodulation with phase unwrapping to address discontinuities \cite{park2007arctangent, alizadeh2019remote}. 

4. Vital Signs Estimation: This stage typically utilizes the discrete Fourier transform (DFT) spectrum to leverage the periodicity and distinct frequency bands of heartbeat and respiration \cite{sacco2020fmcw,alizadeh2019remote,antolinos2020cardiopulmonary}. However, the DFT resolution is inherently limited by its frequency grid, leading to low-resolution estimates and difficulties in resolving interfering harmonic components \cite{paterniani2023radar}. Phase regression \cite{adib2015smart}, may address this limitation by leveraging dominant tonal components, while orthogonal projection techniques, such as the notch filter in \cite{xiong2022vital}, enhance HR estimation by suppressing high-order respiration harmonics. Despite these advancements, both approaches are constrained by their reliance on a small set of frequencies within the DFT spectrum, limiting their ability to fully separate overlapping HR and RR harmonics.

While human trials are critical for validating radar systems and algorithms, their implementation is often hindered by logistical, ethical, and resource constraints, slowing the deployment of these technologies in healthcare and IoT applications. Phantoms present a valuable solution, acting as controlled intermediaries for evaluating radar performance and optimizing algorithmic parameters prior to monitoring humans. Several studies have utilized phantoms to simulate thoracic motions, including a mockup structure with a vibration exciter \cite{bakhtiari2011compact}, a robotic phantom driven by a linear actuator \cite{islam2019programmable}, and a gelatin-coated metal plate \cite{marty2023investigation}. However, these designs are often limited to single-person scenarios and rely on oscillations with predetermined frequencies and amplitudes, lacking the complexity required to emulate real-world multi-person NCVSM in cluttered environments. 

This paper introduces two key contributions toward robust multi-person localization and NCVSM using MIMO FMCW radar. First, we developed a custom hardware phantom designed to simulate multi-person NCVSM in realistic, cluttered environments, utilizing recorded impedance data from monitored humans \cite{schellenberger2020dataset}. The phantom's adaptable design enables repeatable and rapid validation of radar systems and algorithms under diverse conditions, addressing healthcare provider requirements through IoT connectivity. This validation framework played a pivotal role in refining the algorithm and determining optimal configurations for the effective monitoring of multiple individuals in real-world, cluttered environments.

Second, we designed an algorithm that addresses two critical challenges: 1. Multi-person localization in crowded settings, achieved by exploiting the joint sparse representation of individuals through a dedicated signal model informed by cardiopulmonary properties, and 2. harmonics-resilient vital signs estimation, extending the Vital Signs-based Dictionary Recovery (VSDR) method in \cite{eder2023sparsity}, termed E-VSDR, to mitigate interfering respiration harmonics. This method utilizes harmonics-free dictionaries tailored to cardiopulmonary activity to accurately recover the vital signs even in the presence of overlapping harmonics. Additionally, a signal refinement procedure is introduced to enhance accuracy by leveraging the continuity of vital sign estimates through a real-time averaging scheme and adaptively adjusting the frequency bands based on the estimated outcomes.

This study demonstrated superior performance compared to state-of-the-art techniques in multi-person localization \cite{ahmad2018vital,gao2019experiments,han2021detection,konno2014experimental,kim2020low,xiong2022vital,wang2021multi,wang2021driver} and vital signs estimation \cite{adib2015smart,sacco2020fmcw,alizadeh2019remote,antolinos2020cardiopulmonary,xiong2022vital}. Validation was conducted through $12$ phantom trials and $12$ human trials - $9$ single-person and $3$ multi-person trials involving $3$ participants each. In the localization analysis, the proposed method was the only approach to successfully detect and accurately position all subjects in the tested scenarios. In the NCVSM analysis, the proposed E-VSDR method consistently outperformed competitors, even when they were augmented with the suggested refinement procedure, highlighting the strength of the core harmonics-resilient dictionary-based approach. Specifically, for multi-person human trials, the E-VSDR method achieved an average of $94.14\%$, $98.12\%$ and $98.69\%$ of RR estimation errors within $2$, $3$, and $4$ breaths per minute, respectively. For HR estimation, it attained $87.10\%$, $94.12\%$ and $95.54\%$ within the same thresholds of beats per minute.

The rest of this paper is organized as follows: Section \ref{sec:Model} introduces the radar signal model and problem formulation, applicable for both SIMO and MIMO ULA setups. Section \ref{sec:Proposed_algorithm} details the proposed methodology for robust multi-person localization and NCVSM. Section \ref{sec:phantom} describes the phantom validation process and its components. Section \ref{sec:Performance} presents the performance evaluation, beginning with phantom trials and concluding with human trials. Finally, Section \ref{sec:Conclusion} summarizes the findings of this study.

\vspace{-0.15cm}
\section{Signal Model and Problem Formulation}
\label{sec:Model}
This section presents an FMCW signal model and problem formulation applicable for both SIMO and MIMO ULA setups, relying on derivations from our previous work \cite{eder2023sparsity,eder2023sparse}. Building on this model, in Section \ref{sec:Proposed_algorithm} we detail the proposed solution for robust multi-person localization and vital signs monitoring for real-world, cluttered environments.

\vspace{-0.3cm}
\subsection{Signal Model}
\begin{figure}[t!]
    \centering
    \hspace{-0.1cm}\includegraphics[width=0.49\textwidth]{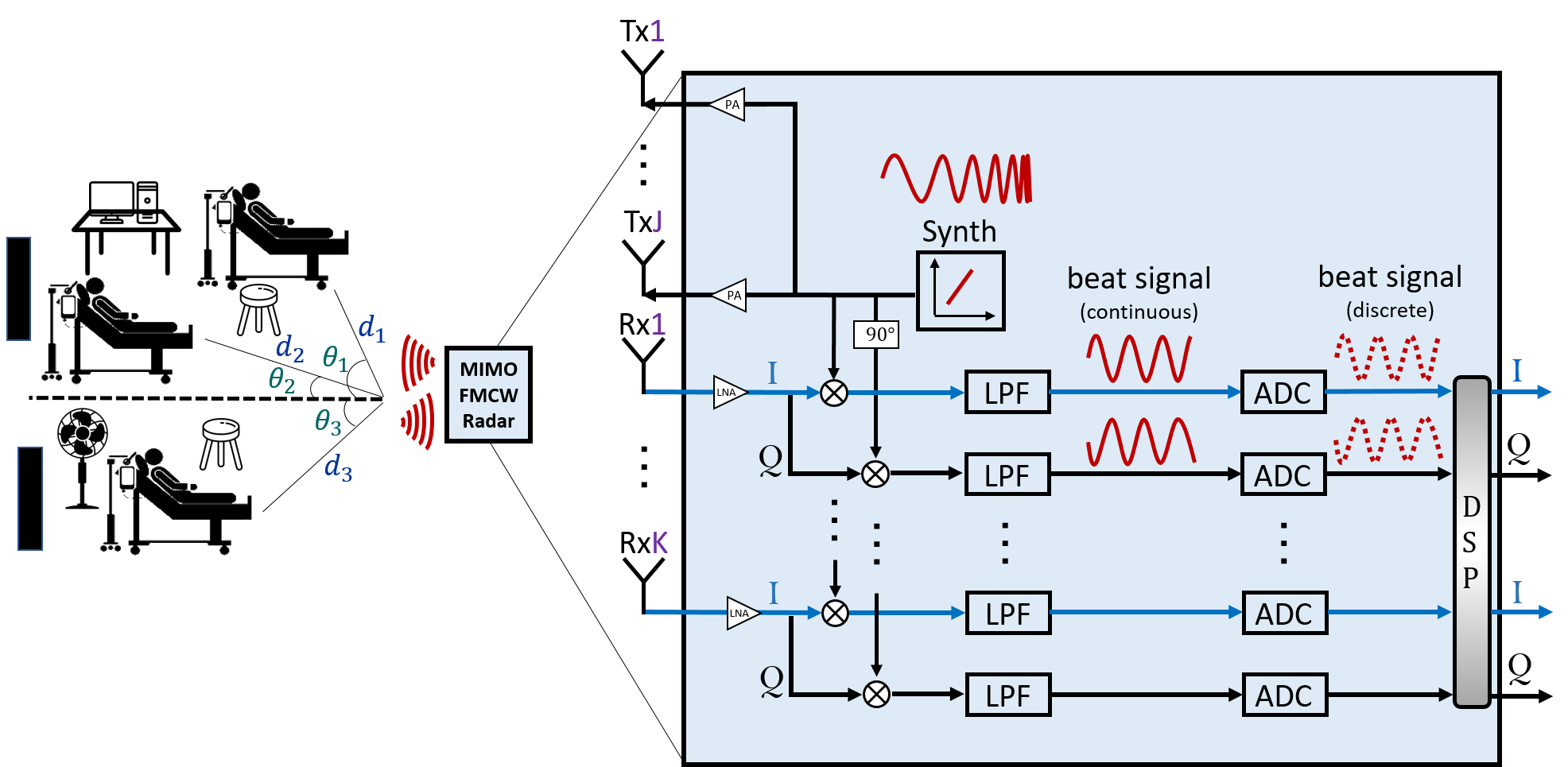}\vspace{0.1cm}
    \caption{A schematic illustration of the main components of a MIMO ULA FMCW radar and a multi-object scenario, with objects located at varying angles and radial distances from the radar.\vspace{-0.5cm}}
    \label{fig:FMCW_radar_MP}
\end{figure}

Consider an FMCW radar in a MIMO ULA setup containing $J \geq1 $ transmitters and $K \geq 1$ receivers, as illustrated in $\textrm{Fig.}\hspace{0.1cm}$\ref{fig:FMCW_radar_MP}. The $j$'th transmitter emits $L$ frames of \textit{chirp} signals \cite{iovescu2017fundamentals} at a frame rate of $f_s$, designed to localize and monitor the vital signs of $Z\geq 1$ humans within a cluttered environment containing $U \geq Z$ objects distributed across various angular and radial positions relative to the radar antennas. The $k$'th receiver captures the reflected echoes, which are separated into the I/Q channels, mixed with the transmitted signal, and sampled by an ADC with interval $T_f$ to produce discrete baseband (\textit{beat}) signals of length $N$ \cite{alizadeh2019remote, turppa2020vital}. The amplitudes of these signals are attenuated based on the radar cross-section (RCS) of the reflecting objects. 

For a single transmitter and $K$ receivers located at ${r_k} \triangleq \left( {k - 1} \right)\lambda/2$, $k = 1, ..., K$, where ${{{\lambda}}}$ is the \textit{chirp}'s maximal wavelength, the $3$D  SIMO FMCW \textit{beat} signal model representing $U$ objects in the radar's FOV can be expressed as \cite{eder2023sparse}:
\begin{equation}
    \label{U_objects_model}
{y\left[ {n,k,l} \right] = \sum\limits_{u = 1}^U {{x_u}{e^{j\left( {2\pi {f_u}n{T_f} + \frac{{2\pi }}{\lambda }{r_k}\sin {\theta _u} + {\psi _u}\left[ l \right]} \right)}}}  + w\left[ {n,k,l} \right]},
\end{equation}
for $n=1,...,N$ \textit{fast-time} samples, $k=1,...,K$ receivers and $l=1,...,L$ \textit{slow-time} frames. Here, $ \{w[n,k,l]\}$ is a $3$D sequence of zero mean i.i.d. complex Gaussian noise with variance ${\sigma}^2$. The received signal is comprised of $U$ components where the $u$'th component is characterized by four parameters: 1. a constant amplitude $x_u$, related to the RCS of the $u$'th object. 2. a \textit{beat} frequency $f_u$ which is proportional to the $u$'th object's radial distance $d_u$ by $f_u\triangleq \frac{2S}{c}d_u$, where $c$ is the speed of light and $S \triangleq B/T_c$ corresponds to the rate of the frequency sweep with $B$ and $T_c$ denoting the \textit{chirp}'s total bandwidth and duration, respectively. 3. an azimuth angle $\theta_u$ and 4. a \textit{slow-time} varying phase term $\psi_u\left[l\right]$ that tracks small vibrations of the $u$'th object.

Assuming that each object has a distinct pair of distance and angle $\{d_u,\theta_u\}$, we can rewrite the model in (\ref{U_objects_model}) for $M\geq U$ general radial distances $\{d_m\}_{m=1}^M$ and $P \geq U$ general azimuth angles $\{\theta_p\}_{p=1}^P$ as
\begin{equation}
\label{MP_objects_model}
y\left[ {n,k,l} \right] = \sum\limits_{p = 1}^P {\sum\limits_{m = 1}^M {{x_{m,p}}{e^{j\left( { \omega_m\left[n\right] + {\phi _p}\left[ k \right]} + {\psi _{m,p}}\left[ l \right] \right)}}} }  + w\left[ {n,k,l} \right],
\end{equation}
where each $\{m,p\}$ component is associated with reflection from a different distance-angle pair $\{d_m,\theta_m\}$, including reflections from $Z $ monitored persons. Based on the latter, $x_{m,p}$ denotes the \textit{beat} amplitude of the $\{m,p\}$'th component, which can be zero if there is no reflection and includes the unknown amplitudes related to the RCS of the monitored individuals. The \textit{fast-time} modulated function $\omega_m\left[n\right]$ is defined by
\begin{equation}
    \label{omega_m[n]}
\omega_m\left[n\right]\triangleq 2\pi {f_m}n{T_f},\quad n=1,\ldots,N,
\end{equation}
where each \textit{beat} frequency $f_m$ is distinct and proportional to a different radial distance from the radar $d_m$ by
\begin{equation}
    \label{fm_dm}
{f_m} \triangleq \frac{{2S}}{c}{d_m},\quad m=1,\ldots,M\leq N,
\end{equation}
where the $Z$ unknown human distances $\{d^{(z)}\}_{z=1}^Z$ are included in the distances $\{d_m\}_{m=1}^M$. The azimuth angles $\{\theta_p\}_{p=1}^P$ are reflected in the following phase shifts due to the ULA antenna geometry:
\begin{equation}
\label{antenna_phase}
{\phi _p}\left[ k \right] = \frac{{2\pi }}{\lambda }{r_k}\sin {\theta _p},\quad k = 1, \ldots ,K \leq P,
\end{equation}
where the $Z$ unknown human angles $\{\theta^{(z)}\}_{z=1}^Z$ are among the angles $\{\theta_p\}_{p=1}^P$. Finally, the \textit{slow-time} varying term ${\psi _{m,p}}\left[ l \right]$ of each component is given by
\begin{equation}
    \label{psi_l_human}
{\psi _{m,p}}\left[ l \right] \triangleq \frac{{4\pi }}{\lambda }\left( {{d_m} + {v_{m,p}}\left[ l \right]} \right),\quad l = 1, \ldots ,L,
\end{equation}
where the vibration function $v_{m,p}\left[ l \right]$  is generally modeled for both human and clutter objects by
\begin{equation}
\label{vib_func} 
{v_{m,p}}\left[ l \right] \triangleq \sum\limits_{q = 1}^Q {a_{m,p}^{\left( q \right)}\cos \left( {2\pi g_{m,p}^{\left( q \right)}l{T_s}} \right)} ,\quad l = 1, \ldots ,L\geq Q.
\end{equation}
The pairs $\{a_{m,p}^{\left( q \right)},g_{m,p}^{\left( q \right)}\}_{q=1}^Q$ are the corresponding amplitudes and frequencies, with the latter being limited by the \textit{slow-time} frame rate $f_s\triangleq 1/T_s$ according to $\{ g_{m,p}^{\left( q \right)} \}_{q=1}^Q \in \left[0\hspace{0.2cm}f_s/2\right)$ for each $\{m,p\}$ component. The generalized vibration model in (\ref{vib_func}) enables effective representation of both static and vibrating objects, including harmonic components, through suitable choices of  $\{ a_{m,p}^{\left( q \right)} \}_{q=1}^Q$ and $\{ g_{m,p}^{\left( q \right)} \}_{q=1}^Q$. 

In this work, we investigate multi-person NCVSM of an unknown number of people $Z$. Their thoracic vibrations, which are included in $\{{v_{m,p}}\left[ l \right]\}$, are denoted by $\{{v^{\left( z \right)}}\left[ l \right]\}_{z=1}^Z$, and satisfy
\begin{equation}
    \label{vib_func_humans}
{v^{\left( z \right)}}\left[ l \right] \triangleq \sum\limits_{q = 1}^Q {a_q^{\left( z \right)}\cos \left( {2\pi g_q^{\left( z \right)}l{T_s}} \right)} ,\quad l = 1, \ldots ,L,
\end{equation}
with amplitudes $\{a_q^{\left( z \right)}\}_{q=1}^Q$ and a frequency set $\{g_q^{\left( z \right)}\}_{q=1}^Q$ that includes the unknown $Z$ pairs of HR and RR, denoted by $\left\{ {f_H^{\left( z \right)},f_R^{\left( z \right)}} \right\}_{z = 1}^Z$.

\vspace{-0.2cm}
\subsection{Problem Formulation}
\begin{figure}[htbp!]  
\begin{center}
\hspace{-0.3cm}
\subfigure{\label{}\includegraphics[width=0.49\textwidth]{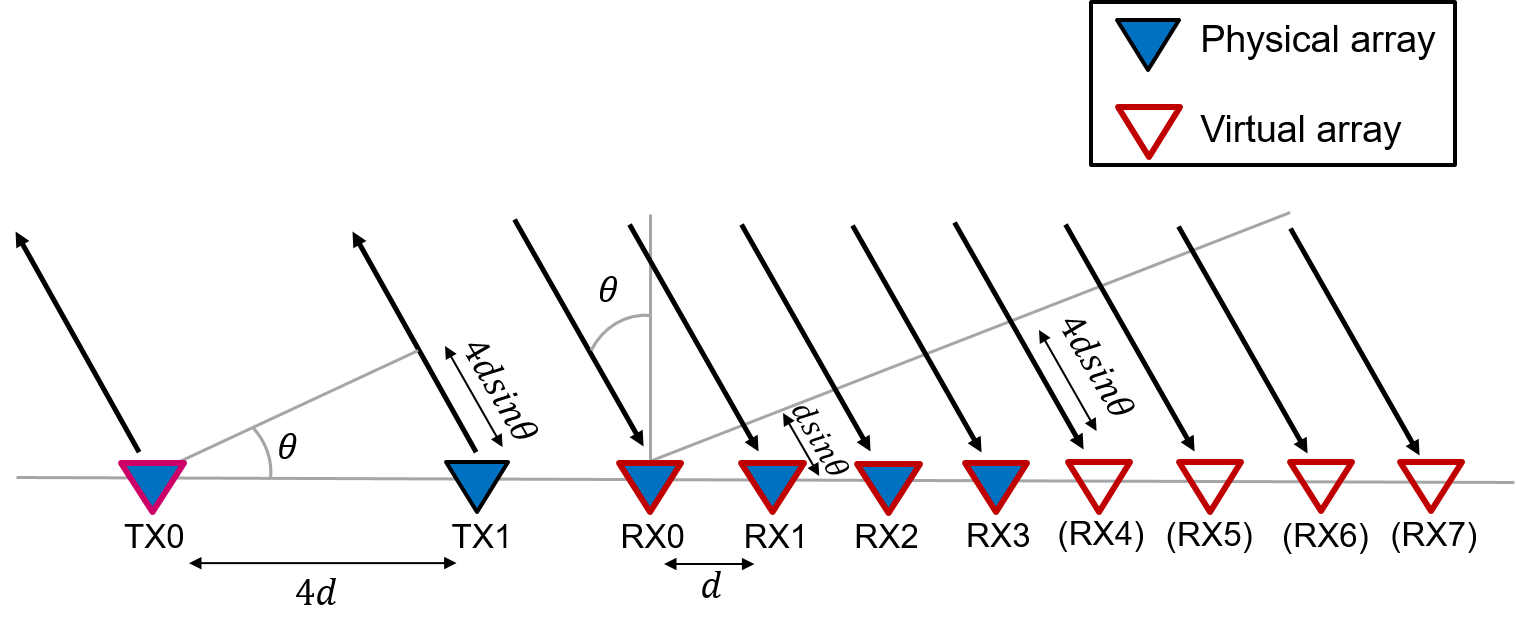}}
\end{center} 
\caption{An illustration of the outcome of a TDM technique, which creates a virtual $1\times{8}$ SIMO ULA given a physical $2\times{4}$ MIMO ULA.}   
\label{fig:MIMO_TDM}
\vspace{-0.4cm}
\end{figure}

Based on the signal model presented above, the following subsection introduces the problem formulation through a simplified matrix representation to facilitate the analysis. Specifically, by defining the complex amplitudes
\begin{equation}
    \label{x_m_p}
{\tilde x_{m,p}}\left[ l \right] \triangleq {x_{m,p}}{e^{j{\psi _{m,p}}\left[ l \right]}},\quad l = 1,\ldots,L,
\end{equation}
the received samples from (\ref{MP_objects_model}) can be arranged into the matrix ${\bf{Y}}_l \in \mathbb{C}^{N\times{K}}$ for each frame. This leads to the model
\begin{equation}
\label{Y=AXB+W}
{\bf{Y}}_l = {\bf{A}}{\bf{X}}_l{{\bf{B}}} + {\bf{W}}_l,\quad l = 1,\ldots,L,
\end{equation}
where ${\bf{A}}\in \mathbb{C}^{N\times M}$ is a known range-related Vandermonde matrix, whose entries are given by ${\bf{A}}\left( {n,m} \right) \triangleq {e^{j{\omega_m}\left[ n \right]}}$ (\ref{omega_m[n]}), ${\bf{B}} \in \mathbb{C}{^{P \times K}}$ is a known angle-related matrix whose entries are given by ${\bf{B}}\left( {p,k} \right) \triangleq {e^{j{\phi _p}\left[ k \right]}}$ (\ref{antenna_phase}), ${{{\bf{X}} }_l} \in \mathbb{C}{^{M \times P}}$ is an unknown matrix of complex amplitudes where ${{{\bf{X}} }_l}\left( {m,p} \right)  \triangleq {\tilde x_{m,p}}\left[ l \right] $ (\ref{x_m_p}),  and ${{\bf{W}}_l} \in \mathbb{C}{^{N \times K}}$ is the noise matrix where ${{{\bf{W}}}_l}\left( {n,k} \right) \triangleq w \left[ n,k,l \right]$ (\ref{U_objects_model}). We note that while the model in (\ref{Y=AXB+W}) is designed for a SIMO ULA setup, it is also applicable to a MIMO ULA by employing orthogonal transmission schemes, such as TDM \cite{li2021signal}. This approach effectively creates a virtual SIMO ULA with a number of virtual receivers equal to the product of the transmitters with the receivers, as illustrated in Fig. \ref{fig:MIMO_TDM}, thereby significantly enhancing angular resolution.

To enable continuous vital signs monitoring, the FMCW radar periodically transmits, receives, and processes frames of chirp signals throughout the monitoring session. At each predefined time interval \( T_{\textrm{int}} \), the sequence \(\{{\bf{Y}}_l\}_{l=1}^L\) in (\ref{Y=AXB+W}) is constructed by aggregating all \( L \) frames recorded within the preceding time window \( T_{\textrm{win}} \), where \( L \) is given by \( L = T_{\textrm{win}} f_s \). The data model in (\ref{Y=AXB+W}) is assumed to exhibit the following properties:  
\begin{enumerate}[{A}-1]  
\item \label{A-1} The monitored individuals remain stationary, with only slight thoracic movements due to cardiopulmonary activity. As a result, the \(\{m,p\}\) coordinates in \(\{{\bf{X}}_l\}_{l=1}^L\) corresponding to their locations \(\{d^{(z)},\theta^{(z)}\}_{z=1}^Z\) (\ref{fm_dm}), (\ref{antenna_phase}), are fixed and joint across all \(L\) frames.  
\item \label{A-2} The number of objects within the radar's FOV, $U$, satisfies \(U \ll MP\), meaning that \(\{{\bf{X}}_l\}_{l=1}^L\) are \(U\)-sparse matrices.  
\end{enumerate}

Based on the model in (\ref{Y=AXB+W}), the first goal is to estimate the number of individuals, \( Z \), along with their spatial locations \(\{d^{(z)},\theta^{(z)}\}_{z=1}^Z\) (\ref{fm_dm}), (\ref{antenna_phase}), by recovering \(\{{\bf{X}}_l\}_{l=1}^L\) and identifying the corresponding \(\{m,p\}\) indices. Subsequently, the second objective is to continuously monitor each detected individual's HR and RR by extracting their thoracic vibrations $\{{v^{\left( z \right)}}\left[ l \right]\}_{z=1}^Z$ (\ref{vib_func_humans}) encoded in \(\{{\bf{X}}_l\}_{l=1}^L\), and estimating the vital pairs \(\{ f_H^{(z)}, f_R^{(z)} \}_{z=1}^Z\) at each \( T_{\textrm{int}} \). 


\vspace{-0.2cm}

\section{Robust Multi-Person Localization and Vital Signs Monitoring}
\label{sec:Proposed_algorithm}
Aided by the designed phantom detailed in Section \ref{sec:phantom}, we developed a robust algorithm for multi-person localization and vital signs monitoring in real-world, cluttered environments, employing an FMCW radar in a MIMO ULA setup. Below, we detail each stage of the proposed approach, based on the model presented in Section \ref{sec:Model}.

\begin{figure*}[htbp!]  
\begin{center}
\hspace{-0.0cm}
\subfigure{\label{}\includegraphics[width=1\textwidth]{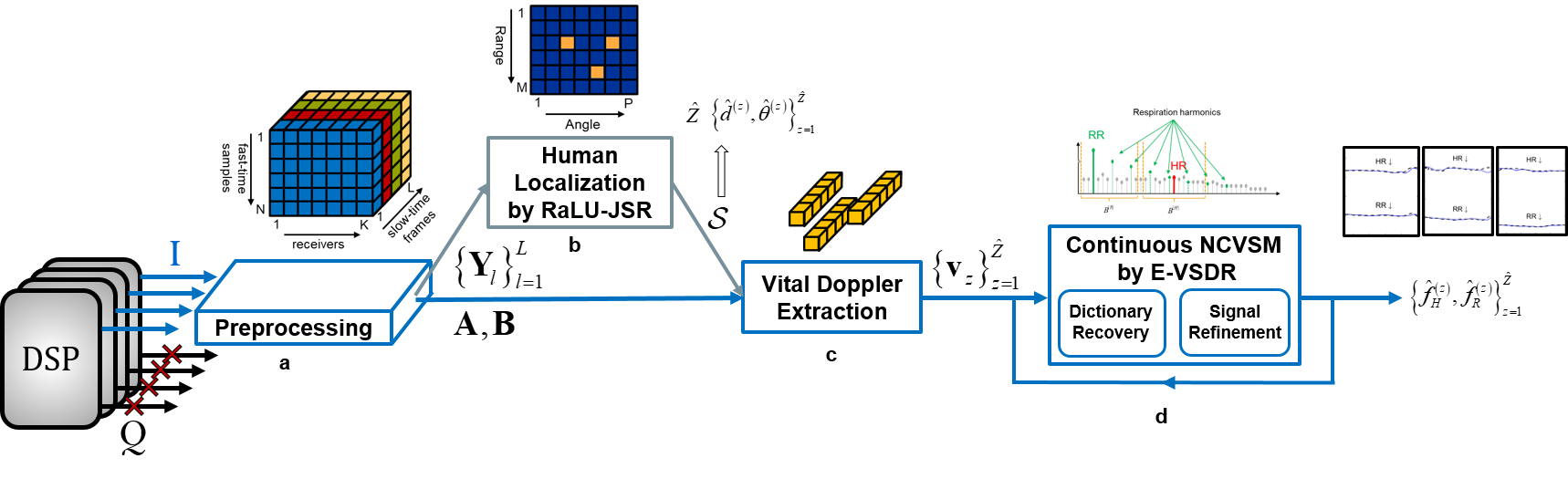}}
\end{center} 
\vspace{-0.5cm}
\caption{Block diagram of the proposed algorithm for robust multi-person localization and vital signs monitoring using SIMO or MIMO ULA FMCW Radar.}   
\label{fig:block_diagram}
\vspace{-0.3cm}
\end{figure*}

 \subsection{Preprocessing}
\label{Pre_Proc}
The first step of each monitoring iteration involves preliminary processing of the radar channel data to construct the $3$D cube $\{{\bf{Y}}_l\}_{l=1}^L$ that satisfies the model in (\ref{Y=AXB+W}) with a high SNR (Fig. \ref{fig:block_diagram}.a). The subsequent step entails assembling the known matrices $\bf{A}$ and $\bf{B}$ based on predefined frequency and angle grids.

Recall that each element of $\{{\bf{W}}_l\}_{l=1}^L$ (\ref{Y=AXB+W}) comes from a zero-mean i.i.d. Gaussian distribution. Hence, similarly to \cite{eder2023sparsity} we reduce the noise variance by utilizing the slowness of thoracic movement relative to the frame period $T_s$ \cite{alizadeh2019remote}. To this end, we transmit $G>1$ consecutive chirps at each frame and coherently combine them to produce a frame with a single chirp that satisfies (\ref{Y=AXB+W}) with variance smaller by a factor of $G$, w.r.t. transmission of only a single chirp per frame.

Next, we set the dictionary matrix ${\bf{A}}$ by assuming that the distance-related frequencies $\{f_m\}_{m=1}^M$ (\ref{fm_dm}) lie on the Nyquist grid, i.e.,
\begin{equation}
    \label{fm_Nyquist_grid}
f_m  =\frac{{{f_{\textrm{ADC}}}}}{N} {i_m}{\rm{,}}\quad {\rm{ }}{i_m} = 0,\ldots,M - 1,\vspace{-0.1cm}
\end{equation}
where $f_{\textrm{ADC}}\triangleq1/T_f$ is determined by the ADC component. We note using (\ref{fm_dm}) and (\ref{fm_Nyquist_grid}), that the maximal detectable distance is $d_{\textrm{max}}=\frac{cf_{\textrm{ADC}}}{{2SN}}\left(M-1\right)$, and the range resolution is $d_{\textrm{res}}=\frac{cf_{\textrm{ADC}}}{{2SN}}=\frac{c}{2B}$ since $N=f_{\textrm{ADC}}T_c$ and $ST_c=B$. Using (\ref{fm_Nyquist_grid}) and since ${\bf{A}}\left( {n,m} \right) \triangleq {e^{j2\pi {f_m}n{T_f}}}$, we have that ${\bf{A}}\left( {n,m} \right) = {e^{j2\pi \frac{i_m}{N}n}}$. As for the dictionary matrix ${\bf{B}}$, the phase shifts $\{{\phi _p}\left[ k \right] \}_{k=1}^K$ (\ref{antenna_phase}) are set to cover a FOV of $180$ [$^{\circ}$] according to the following angle grid
\vspace{-0.2cm}
\begin{equation}
\label{theta_p_grid}
{\theta _p} =  - 90 + {i_p}{\Delta _\theta },\quad {i_p} = 0, \ldots ,P - 1\quad ,P = \frac{{180}}{{{\Delta _\theta }}},
\end{equation}
where ${\Delta _\theta }$ denotes the spacing of the angle grid. Then, by (\ref{antenna_phase}), (\ref{theta_p_grid}) and since ${\bf{B}}\left( {p,k} \right) \triangleq {e^{j{\phi _p}\left[ k \right]}}$, we have that ${\bf{B}}\left( {k,p} \right)= {e^{j\pi \left( {k - 1} \right)\sin \left( { - 90 + {i_p}{\Delta _\theta }} \right)}}$. \vspace{-0.5cm}

\subsection{Human Localization by RaLU-JSR}
\label{Human_Localization}
To estimate the number of humans $Z$ and their spatial location $\{d^{(z)},\theta^{(z)}\}_{z=1}^Z$ (Fig. \ref{fig:block_diagram}.b), we use the assembled dictionaries $\bf{A}$ and $\bf{B}$ as well as $\{{\bf{Y}}_l\}_{l=1}^L$ for the first $L={T_{\textrm{loc}}}f_s$ frames, where ${T_{\textrm{loc}}}$ denotes the duration of localization. As this study focuses on the continuous monitoring of multiple stationary subjects, we recover $\{{\bf{X}}_l\}_{l=1}^L$ and the corresponding support only once. Here, the support is denoted by $\cal S$ and defined as the set of $2$D $\{m,p\}$ indices whose cardinality corresponds to the number of individuals $Z$ and whose indices point to their respective range-angle locations $\{d^{(z)},\theta^{(z)}\}_{z=1}^Z$ by (\ref{fm_dm}), (\ref{fm_Nyquist_grid}) and (\ref{theta_p_grid}). Our approach removes the need to recover $\{{\bf{X}}_l\}_{l=1}^L$ and $\cal S$ at each monitoring iteration, as further explained in Subsection \ref{Dopp_Rec}.

To this end, we start with clutter suppression utilizing prior knowledge of human-typical pulse and breathing frequencies by extending the vital-based spectral filter introduced in \cite{eder2023sparsity} for a single receiver, here to the case of $K>1$ receivers. We select the data of each $k$'th antenna by defining ${{\bf{Y}}^{\left( k \right)}} \in \mathbb{C}{^{N \times L}}:{{\bf{Y}}^{\left( k \right)}}\left( {n,l} \right) \triangleq {{\bf{Y}}_l}\left( {n,k} \right)$ (\ref{Y=AXB+W}). Then, each filtered matrix is given by
\begin{equation}
    \label{Y_k_filtered}
{{{\bf{\bar Y}}}^{\left( k \right)}} = \frac{1}{L}{\left( {{\bf{F}}_L^H\left( {{\bf{\Pi }} \odot {{\bf{F}}_L}{{\bf{Y}}^{\left( k \right)T}}} \right)} \right)^T},\quad k = 1, \ldots ,K
\end{equation}
where ${{\bf{F}}_L}$ is a full $L$-size DFT matrix, ${\bf{\Pi }}\in \mathbb{R}^{L}$ denotes a window function corresponding to the vital frequencies in ${B^{\left( H \right)}} \cup {B^{\left( R \right)}}$ with $B^{(R)}$ and $B^{(H)}$ respectively denoting the frequency bands of respiration and heartbeat at rest, and $\odot$ denotes the element-wise product.
We then reshape the filtered data back to the original structure as in (\ref{Y=AXB+W}) denoted by $\{{\bf{\bar{Y}}}_l\}_{l=1}^L$. 
 
Next, by assumptions \ref{A-1} and \ref{A-2}, the $U$-sparse matrices $\left\{ {{{\bf{X}}_l}} \right\}_{l = 1}^L$ share joint support. Hence, we propose to recover them from $\{{\bf{\bar{Y}}}_l\}_{l=1}^L$ by promoting the joint sparsity via the following $3$D $l_{2,1}$-norm regularized Least-Squares (LS) \cite{ruppert1994multivariate} problem given ${\bf{A}}$ and ${\bf{B}}$:
\begin{equation}
    \label{l_{2,1}_LS}
\hat{\mathbb{X}}= \mathop {\arg \min }\limits_{\mathbb{X} \in \mathbb{C}^{M\times{P}\times{L}}} \frac{1}{{2L}}\sum\limits_{l = 1}^L {\left\| {{{\bf{\bar{Y}}}_l} - {\bf{A}}{{\bf{X}}_l}{\bf{B}}} \right\|_F^2}  + \gamma {{\left\| \mathbb{X} \right\|}_{2,1}},\vspace{-0.15cm}
\end{equation}
where ${\left\| \mathbb{X} \right\|_{2,1}}$ refers to the sum of all $l_2$ norms over the frame (third) dimension of the $3$D tensor $\mathbb{X} \in \mathbb{C}^{M\times{P}\times{L}}$ which concatenates the frames of ${\left\{ {{{\bf{X}}_l}} \right\}_{l = 1}^L}$ and $\gamma\geq 0$ is the regularization parameter. To solve (\ref{l_{2,1}_LS}) we extend the JSR algorithm proposed in \cite{rossman2019rapid}, which is based on the fast iterative soft-thresholding algorithm (FISTA) \cite{beck2009fast,palomar2010convex}, to accommodate the proposed bilinear formulation (\ref{Y=AXB+W}). The method is called RaLU-JSR: \textbf{Ra}dar \textbf{L}ocalization of h\textbf{U}mans via \textbf{J}oint \textbf{S}parse \textbf{R}ecovery and is given in Algorithm \ref{alg_loc_3D_JSR_FISTA}.

\begin{algorithm}[h!]
\begin{algorithmic}
\caption{RaLU-JSR for minimizing (\ref{l_{2,1}_LS})}
\State \hspace{-0.4cm} \textbf{Input:} $\{{\bf{\bar{Y}}}_l\}_{l=1}^L,{\bf{A}},{\bf{B}},{L_f},\gamma  > 0, I$
\State \hspace{-0.3cm}\textbf{Initialize:} $i = 1,{t^{(1)}} = 1,\left\{ {{\bf{Z}}_l^{\left( 1 \right)} = {\bf{X}}_l^{\left( 0 \right)} = {{\bf{0}}^{M \times P}}} \right\}_{l=1}^L$
\State \hspace{-0.1cm}{\bf{while}}{\rm{ }}$i < I_{\rm{max}}${\rm{ or stopping criteria not fulfilled }}{\bf{do}}
\State \hspace{-0.2cm} \textbf{1:} ${\bf{G}}_l^{\left( i \right)} = {\bf{Z}}_l^{\left( i \right)} - \frac{1}{{{L_f}}}{{\bf{A}}^H}\left( {{\bf{AZ}}_l^{\left( i \right)}{\bf{B}} - {{\bf{Y}}_l}} \right){{\bf{B}}^H}$, $l = 1,...,L$
\State \hspace{-0.2cm} \textbf{2:} $\left\{ {{\bf{X}}_l^{\left( {i + 1} \right)}} \right\}_{l = 1}^L = {\cal T}_{\frac{\gamma }{L_f}}^{\left( {3D} \right)}\left( {\left\{ {{\bf{G}}_l^{\left( i \right)}} \right\}_{l = 1}^L} \right){\rm{ }}$
\State \hspace{-0.2cm} \textbf{3:} ${{\rm{t}}^{\left( {i + 1} \right)}} = 0.5\left( {1 + \sqrt {1 + 4{t^{\left( i \right)2}}} } \right)$
\State \hspace{-0.2cm} \textbf{4:} ${\bf{Z}}_l^{\left( {i + 1} \right)} = {\bf{X}}_l^{\left( i \right)} + \frac{{{t^{\left( i \right)}} - 1}}{{{{\rm{t}}^{\left( {i + 1} \right)}}}}\left( {{\bf{X}}_l^{\left( i \right)} - {\bf{X}}_l^{\left( {i - 1} \right)}} \right),{\rm{ }}l = 1,...,L$
\State \hspace{-0.2cm} \textbf{5:} $i \leftarrow i + 1$
\State \hspace{-0.1cm}{\bf{end while}}

\State \hspace{-0.4cm}\textbf{Return:} $\left\{ {{\bf{X}}_l^{\left( {I} \right)}} \right\}_{l = 1}^L$
\label{alg_loc_3D_JSR_FISTA}
\end{algorithmic}
\end{algorithm}


In Algorithm \ref{alg_loc_3D_JSR_FISTA}, ${I}$ denotes the maximal number of iterations, $L_f$ denotes the Lipschitz constant \cite{beck2009fast}. ${L_f} = {\lambda _{\max }}\left( {{{\bf{A}}^H}{\bf{A}}} \right){\lambda _{\max }}\left( {{{\bf{B}}^H}{\bf{B}}} \right)$ with ${\lambda _{\max }}$ denoting the largest singular value. ${\cal T}_\alpha ^{\left( {3D} \right)}\left(  \cdot  \right)$ is the $3$D soft-threshold operator such that for each $\{m,p\}$'th element of ${{\bf{X}}_l^{\left(i+1\right)}}$, $l=1,...,L$, we have
\vspace{-0.2cm}
\begin{equation}
    \label{3D_soft_thr}
    {{\bf{X}}_l^{\left(i+1\right)}}\left( {m,p} \right) = \max \left( {0,1 - \alpha / {\|\mathbb{G}^{\left( i \right)}\left( {m,p} \right)\|_2 } } \right) {\bf{G}}_l^{\left( i \right)}\left( {m,p} \right),
\end{equation}
where the tensor $\mathbb{G}^{\left( i \right)} \in \mathbb{C}^{M\times{P}\times{L}}$ concatenates all $L$ matrices ${ {{\bf{G}}_l^{\left( i \right)}}}\in \mathbb{C}^{M\times{P}}$, $l=1,...,L$, as detailed in Algorithm \ref{alg_loc_3D_JSR_FISTA}.

The support $\mathcal{S}$ is estimated by first taking the average power across the frames of $\{ {{\bf{X}}_l^{\left( {{I}} \right)}} \}_{l = 1}^L$ which results in a $2$D range-angle localization map, denoted by ${\bf{\bar X}} \in \mathbb{R} {^{M \times P}}$. Then, various methods can be used to determine $\mathcal{S}$ from ${\bf{\bar X}}$, (and corrspondingly $\hat{Z}$, and $\{\hat{d}^{(z)},\hat{\theta}^{(z)}\}_{z=1}^{\hat{Z}}$), such as CA-CFAR \cite{jalil2016analysis} or $2$D peak-detection \cite{eder2023sparse}. We note that predetermined boundaries of range and angle can serve as a focused region of interest (ROI) in ${\bf{\bar X}}$ for evaluating $\mathcal{S}$.

\vspace{-0.2cm}
\subsection{Vital Doppler Extraction}
\label{Dopp_Rec}
The support evaluated in the localization step allows us to efficiently recover only the Doppler samples related to human vital signs concealed in the input set $\{{\bf{Y}}_l\}_{l=1}^L$ (\ref{Y=AXB+W}) of each monitoring iteration (Fig. \ref{fig:block_diagram}.c). Particularly, using $\mathcal{S}$ we estimate the complex amplitudes $\{{\tilde x_{m,p}}\left[ l \right]\}_{l=1}^L$ (\ref{x_m_p}) of each $z$'th human, denoted by $\{{x_{S\left( z \right)}}\left[ l \right]\}_{l=1}^L$, using the following beamformer over the support:
\begin{equation}
    \label{x_s_z}
{\hat x_{S\left( z \right)}}\left[ l \right] = \frac{1}{{NK}}{\bf{A}}_{S\left( z \right)}^H{{\bf{Y}}_l}{\bf{B}}_{S\left( z \right)}^H,\quad l = 1, \ldots ,L,
\end{equation}
for each human $z=1,...,\hat{Z}$, where ${{\bf{A}}_{S\left( z \right)}} \in \mathbb{C}{^{N \times{1}}}$ and ${{\bf{B}}_{S\left( z \right)}} \in \mathbb{C}{^{1 \times{K}}}$ respectively denote the atoms of $\bf{A}$ and $\bf{B}$ corresponding the $z$'th $2$D index of $\cal{S}$. We note that since ${\bf{A}}\left( {n,m} \right) = {e^{j2\pi \frac{i_m}{N}n}}$, ${\bf{A}}_{S\left( z \right)}^H$ equals to the row of a partial DFT matrix corresponding to the \textit{fast-time} frequency $f_m$ (\ref{fm_Nyquist_grid}) for $m \in {\cal{S}}(z)$. Hence, analogous to \cite{eder2023sparsity}, by selecting $M=N/2$ in (\ref{fm_Nyquist_grid}) (i.e. considering only non-negative \textit{fast-time} frequencies), the estimator in (\ref{x_s_z}) given $\{{\bf{Y}}_l\}_{l=1}^L$ (\ref{Y=AXB+W}) assembled by both the I and Q channels, is comparable to that obtained by simply using a single channel (I or Q), up to a constant factor. To minimize hardware overload and possible concerns of using two channels concurrently, we assume $M=N/2$ and use only the \textit{In-Phase} channel to construct (\ref{Y=AXB+W}). 

Next, since ${\hat x_{S\left( z \right)}}\left[ l \right]$ (\ref{x_s_z}) estimates the phasor terms ${x_{m,p}}{e^{j{\psi _{m,p}}\left[ l \right]}}$ (\ref{x_m_p}) for $\{m,p\} \in \cal{S}$, that modulate the thoracic vibrations $\{{v^{\left( z \right)}}\left[ l \right]\}_{z=1}^Z$ (\ref{psi_l_human})-(\ref{vib_func_humans}), we extract the phase of each $z$'th human using an arctangent demodulation-based method similarly to \cite{alizadeh2019remote,eder2023sparsity,eder2023sparse}:
\begin{equation}
     \label{v_z_est}
{\hat v^{\left( z \right)}}\left[ l \right] = unwrap\left( {\angle \left( {{{\hat x}_{S\left( z \right)}}\left[ l \right]} \right)} \right),\quad l = 1, \ldots ,L,
\end{equation}
where $unwrap\left(\cdot\right)$ denotes the unwrapping procedure described in \cite{alizadeh2019remote}, used since the unambiguous phase range is limited by $(-\pi\hspace{0.1cm}\pi]$ and the angle extraction operator $\angle\left(\cdot\right)$ is based on the four quadrant arctangent function. For convenient analysis, the estimated samples are then concatenated into the vector $\hat{{\bf{v}}}_z \in \mathbb{R}^{L}$ which represents a scaled approximation of $\{{v^{\left( z \right)}}\left[ l \right]\}_{l=1}^L$ in (\ref{vib_func_humans}), for each $z$'th detected human, $z=1,...,\hat{Z}$. \vspace{-0.7cm}

 \subsection{Continuous NCVSM by E-VSDR}
\label{EVSDR}
In the final stage of each monitoring iteration, the vital signs of the detected individuals, $\{ {f_H^{\left( z \right)},f_R^{\left( z \right)}}\}_{z = 1}^Z$, are estimated given the extracted vibrations $\{\hat{{\bf{v}}}_z\}_{z = 1}^{\hat{Z}}$ (\ref{v_z_est}), and recorded for continuous NCVSM (Fig. \ref{fig:block_diagram}.d). In the following, we present an extension to the VSDR approach introduced in \cite{eder2023sparsity}, termed E-VSDR, which is tailored for continuous monitoring in low-SNR conditions where interfering harmonics due to the non-pure sinusoidal nature of human thoracic motion pose challenges \cite{xiong2022vital, hsieh2024multiperson}. Furthermore, E-VSDR incorporates a dedicated adaptive signal refinement procedure that leverages the high rate of output estimates to enhance monitoring accuracy, making it particularly effective for continuous NCVSM in real-world, cluttered environments.

\subsubsection{Preliminaries}
First, according to (\ref{vib_func_humans}), and given the non-overlapping nature of respiration and heartbeat frequency bands, the extracted vibrations \(\{\hat{{\bf{v}}}_z\}_{z=1}^{\hat{Z}}\) (\ref{v_z_est}) can be described by 
\begin{equation}
    \label{v_z_equation}
{{\hat{{\bf{v}}}_z} = {{\bf{D}}^{\left( R \right)}}{\bf{a}}_z^{\left( R \right)} + {{\bf{D}}^{\left( H \right)}}{\bf{a}}_z^{\left( H \right)} + {{\bf{n}}_z}},\quad z=1,\ldots,\hat{Z},
\end{equation}
where ${\bf{D}}^{\left( R \right)}\in \mathbb{R}{^{L \times Q_R}}$ and ${\bf{D}}^{\left( H \right)}\in \mathbb{R}{^{L \times Q_H}}$, ${Q_R},{Q_H} < Q$ respectively denote the respiration and heartbeat dictionaries:
\begin{equation}
    \label{D_def}
\begin{array}{l}
{{\bf{D}}^{\left( R \right)}}\left( {l,q} \right) \triangleq \cos \left( {2\pi  g_q^{\left( R \right)}l{T_s}} \right)\hspace{0.2cm}\textrm{and}\\
{{\bf{D}}^{\left( H \right)}}\left( {l,q} \right)\triangleq \cos \left( {2\pi  g_q^{\left( H \right)}l{T_s}} \right),
\end{array}
\end{equation}
where $\{g_q^{\left( R \right)}\}_{q=1}^{Q_R} \in B^{\left( R \right)}$ and $\{g_q^{\left( H \right)}\}_{q=1}^{Q_H} \in B^{\left( H \right)}$ with $B^{\left( R \right)} \cap B^{\left( H \right)} = \emptyset$. The amplitude vectors  ${\bf{a}}_z^{\left( R \right)}\in\mathbb{R}{^{Q_R}}$ and ${\bf{a}}_z^{\left( H \right)}\in\mathbb{R}{^{Q_H}}$ control the frequency pattern of each $z$'th human vibration through ${\bf{D}}^{\left( R \right)}$ and ${\bf{D}}^{\left( H \right)}$. Finally, $\{{\bf{n}}_z\}_{z=1}^{\hat{Z}}$ denote length-$L$ sequences of i.i.d. noise vectors as a result of the non-linear operations up to (\ref{v_z_est}).

\begin{figure}[htbp!]  
\vspace{-0.3cm}
\begin{center}
\hspace{-0.3cm}
\subfigure{\label{}\includegraphics[width=0.38\textwidth]{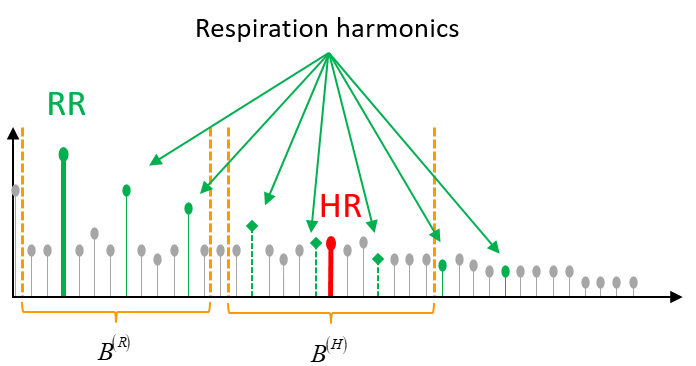}}
\end{center} 
\vspace{-0.3cm}
\caption{Illustration of the spectral components of respiration and heartbeat frequency bands $B^{(R)}$ and $B^{(H)}$, respectively. The overlap between heartbeat components and high-order respiratory harmonics highlights the challenge of accurate HR estimation.}   
\label{fig:RR_harmonics}
\vspace{-0.5cm}
\end{figure}

Next, to mitigate the phenomenon where respiration harmonics may mask the heartbeat tone in \( B^{\left( H \right)} \) \cite{hsieh2024multiperson,paterniani2023radar} (see Fig. \ref{fig:RR_harmonics}), for each \( z \)'th human we divide the frequencies in \( B^{\left( H \right)} \) into two distinct groups: 1. Interfering respiration harmonics, denoted by $\{ {g_q^{({R',z} )}}\}_{q = 1}^{{Q_{R',z}}}$, defined as multiples of the fundamental respiration frequency $f_R^{\left( z \right)}$ that reside in $ B^{\left( H \right)}$:
\begin{equation}
    \label{RR_hamonics_def}
{\left\{ {g_q^{\left( {R',z} \right)} = g_q^{\left( H \right)}:\quad g_q^{\left( H \right)} = {i_q}f_R^{\left( z \right)},\quad {i_q}f_R^{\left( z \right)} \in {B^{\left( H \right)}}} \right\}}.
\end{equation}
2. Non-interfered heart frequencies, denoted by $\{ {g_q^{( {H',z} )}} \}_{q = 1}^{{Q_{H',z}}}$ which are the complementary frequencies to $\{ {g_q^{( {R',z} )}} \}_{q = 1}^{{Q_{R',z}}}$ in $ B^{\left( H \right)}$ that include the true HR, $f_H^{\left( z \right)}$. Consequently, $\{ {g_q^{( {R',z} )}} \}_{q = 1}^{{Q_{R',z}}}$ and $\{ {g_q^{( {H',z} )}} \}_{q = 1}^{{Q_{H',z}}} $ respectively compose the dictionaries ${{\bf{D}}^{( {R',z} )}} \in \mathbb{R}{^{L \times {Q_{R',z}}}}$ and ${{\bf{D}}^{( {H',z} )}} \in \mathbb{R}{^{L \times {Q_{H',z}}}}$ similarly to (\ref{D_def}), where ${{\bf{D}}^{( {R',z} )}} \cup {{\bf{D}}^{( {H',z} )}} =  {{\bf{D}}^{\left( {H} \right)}} $ and ${Q_{H',z}} + {Q_{R',z}}={Q_H}$. We remark that the dictionaries ${{\bf{D}}^{( {R',z} )}} $ and ${{\bf{D}}^{( {H',z} )}} $ are unknown a priori, as they rely on the unknown respiratory fundamental frequency of each individual, $f_R^{\left( z \right)}$. Under these settings, the $z$'th extracted vibration vector ${\hat{{\bf{v}}}_z}$ (\ref{v_z_equation}) can be represented as
\begin{equation}
    \label{v_z_harmonics}
{{\hat{{\bf{v}}}_z} = {{\bf{D}}^{\left( R \right)}}{\bf{a}}_z^{\left( R \right)} + {{\bf{D}}^{\left( {R',z} \right)}}{\bf{a}}_z^{({R'} )} + {{\bf{D}}^{\left( {H',z} \right)}}{\bf{a}}_z^{( {H'} )} + {{\bf{n}}_z}},
\end{equation}
for $z=1,...,\hat{Z}$, where ${{\bf{a}}_z^{( {R'} )}} \in \mathbb{R}^{Q_{R',z}}$ and ${{\bf{a}}_z^{( {H'})}}\in \mathbb{R}^{Q_{H',z}}$ respectively denote the corresponding amplitudes of interfered and non-interfered heartbeat tones. Given the model in (\ref{v_z_harmonics}), and the prominence of respiration and heartbeat tones in cardiopulmonary activity, the amplitude vectors ${{\bf{a}}_z^{( {R} )}}$ and ${{\bf{a}}_z^{( {H'} )}}$ are assumed to be $1$-sparse vectors.

Finally, to allow for short estimation windows but with sufficient frequency resolution, similarly to \cite{eder2023sparsity} we uniformly split the \textit{slow-time} frequency segment $\left[0\hspace{0.2cm}f_s/2\right)$, (which includes $ B^{\left( R \right)}$ and $ B^{\left( H \right)}$) according to a resolution of $1$ bpm. This means that given a dense grid of frequencies
\begin{equation}
    \label{g_q_freqs}
g_q = {h_q}\frac{{{f_s}}}{Q}{\rm{,}}\quad {\rm{ }}{h_q} = 0,...,\frac{Q}{2} - 1,\quad Q=60f_s,
\end{equation}
the frequencies $\{ {g_q^{\left( {R} \right)}} \}_{q = 1}^{{Q_{R}}} $ and $\{ {g_q^{\left( {H} \right)}} \}_{q = 1}^{{Q_{H}}} $ of (\ref{D_def}), constitute subsets of (\ref{g_q_freqs}) according to the limits defined by $ B^{\left( R \right)}$ and $ B^{\left( H \right)}$, respectively. These frequencies are then used to assemble ${\bf{D}}^{\left( R \right)}$ and ${\bf{D}}^{\left( H \right)}$ following (\ref{D_def}).

\vspace{0.2cm}
\subsubsection{Vital Signs Estimation}
We first estimate the RR from ${\hat{{\bf{v}}}_z}$ (\ref{v_z_harmonics}) by leveraging the $1$-sparse property of ${{\bf{a}}_z^{( {R} )}}$ according to
\begin{equation}
    \label{S_R}
{\cal{S}}_R^{(z)}= \mathop {\arg \max }\limits_{q=1,\ldots,Q_R} \{| ({{\bf{D}}^{\left( R \right)T}}{\hat{{\bf{v}}}_z})_q| \},
\end{equation}
where ${{\bf{D}}^{\left( R \right)T}}{\hat{{\bf{v}}}_z}\in \mathbb{R}^{Q_R}$ and ${\cal{S}}_R^{(z)}$ denotes the respiration support of the $z$'th human, which selects the $q$'th frequency within $\{g_q^{\left( R \right)}\}_{q=1}^{Q_R}$ as the RR estimate ${\hat f_R^{\left( z \right)}}$. Next, we subtract the influence of respiration by taking the residual vector ${{\hat{{\bf{v}}}}'_z} \in \mathbb{R}^L$ as
\begin{equation}
    \label{v_prime}
    {{\hat{{\bf{v}}}}'_z} = {\hat{{\bf{v}}}_z}-{\bf{d}}_{{\cal{S}}_R}^{(z)}\hat{a}_{{\cal{S}}_R}^{(z)},
\end{equation}
where ${\bf{d}}_{{\cal{S}}_R}^{(z)}\in \mathbb{R}^L$ is the atom of ${{\bf{D}}^{\left( R \right)}}$ corresponding to ${\cal{S}}_R^{(z)}$, and $\hat{a}_{{\cal{S}}_R}^{(z)}={\bf{d}}_{{\cal{S}}_R}^{(z)T}{\hat{{\bf{v}}}_z}/({\bf{d}}_{{\cal{S}}_R}^{(z)T}{\bf{d}}_{{\cal{S}}_R}^{(z)})\in \mathbb{R}$ is the estimated amplitude over ${\cal{S}}_R^{(z)}$.

Then, to mitigate the impact of interfering respiratory harmonics on HR estimation, we use (\ref{RR_hamonics_def}) and ${\hat f_R^{\left( z \right)}}$ (\ref{S_R}) to first estimate the respiration harmonics $\{ {g_q^{({R',z} )}} \}_{q = 1}^{{Q_{R',z}}} $ and the complementary set $\{ {g_q^{( {H',z} )}} \}_{q = 1}^{{Q_{H',z}}}$ by which we assemble ${{\bf{{D}}}^{\left( {R',z} \right)}}$ and ${{\bf{{D}}}^{\left( {H',z} \right)}}$, respectively. We then subtract the respiratory harmonics from ${{\hat{{\bf{v}}}}'_z}$ (\ref{v_prime}) by
\begin{equation}
    \label{v_prime_prime}
    {{\hat{{\bf{v}}}}''_z}={{\hat{{\bf{v}}}}'_z}-{{\bf{D}}^{\left( {R',z} \right)}}\hat{{\bf{a}}}_z^{({R'} )},
\end{equation}
where $\hat{{\bf{a}}}_z^{({R'} )}=\left({{\bf{D}}^{\left( {R',z} \right)T}}{{\bf{D}}^{\left( {R',z} \right)}}\right)^{-1}{{\bf{D}}^{\left( {R',z} \right)T}}{{\hat{{\bf{v}}}}'_z}$ is the LS solution \cite{ruppert1994multivariate} given ${{\hat{{\bf{v}}}}'_z}$ and ${{\bf{D}}^{\left( {R',z} \right)}}$. Finally, we estimate $f_H^{\left( z \right)}$ given $ {{\hat{{\bf{v}}}}''_z}$ (\ref{v_prime_prime}) and ${{\bf{{D}}}^{\left( {H',z} \right)}}$ by leveraging the $1$-sparse property of ${{\bf{a}}_z^{( {H'} )}}$:
\begin{equation}
    \label{S_H}
{\cal{S}}_H^{(z)}= \mathop {\arg \max }\limits_{q=1,\ldots,Q_{H',z}} \{ |({{{\bf{{D}}}^{\left( {H',z} \right)T}}{\hat{{\bf{v}}}''_z}})_q| \},
\end{equation}
where ${{{\bf{{D}}}^{\left( {H',z} \right)T}}{\hat{{\bf{v}}}''_z}} \in \mathbb{R}^{Q_{H',z}}$ and ${\cal{S}}_H^{(z)}$ denotes the heartbeat support of the $z$'th human, which points to the $q$'th frequency within $\{ {g_q^{( {H',z} )}} \}_{q = 1}^{{Q_{H',z}}}$ that reflects the HR estimate, ${\hat f_H^{\left( z \right)}}$.

\vspace{0.3cm}
\subsubsection{Signal Refinement}
\label{signal_refinement}
The continuous NCVSM framework of this work enables leveraging the ongoing stream of estimates $\{{\hat f_H^{\left( z \right)}},{\hat f_R^{\left( z \right)}}\}_{z=1}^Z$ to manage RBMs and potential overlaps between HR and RR harmonics, aiming to enhance overall monitoring accuracy. Consequently, at each $T_{\rm{int}}$ after predefined $T_{\rm{ref}}$ seconds, a three-stage signal refinement procedure is applied for the monitoring iteration, incorporating the current estimates along with prior ones as follows:
$1.$ For the monitoring iteration at $T_{\rm{ref}}$, we replace all estimates with the median value derived from the samples collected up to that time, which can alleviate the fluctuations often observed at the onset of monitoring. For each $T_{\rm{int}}$ following $T_{\rm{ref}}$: $2.$ the vital estimates $\{{\hat f_H^{\left( z \right)}},{\hat f_R^{\left( z \right)}}\}_{z=1}^Z$ are replaced with the average of the estimations acquired in the last $T_{\rm{avg}}^{(H)}$ and $T_{\rm{avg}}^{(R)}$ seconds, respectively. This principle serves as an online filter that smooths the curve of estimates to ensure a gradual rate of changes in the vital signs. $3.$ the fixed bands of respiration and heartbeat, ${B^{\left( R \right)}}$ and ${B^{\left( H \right)}}$ respectively, are replaced with adaptive bands, centered around the vital estimates $\{{\hat f_R^{\left( z \right)}},{\hat f_H^{\left( z \right)}}\}_{z=1}^Z$ with small frequency margins. This adjustment focuses the frequency search within a limited range that more reliably tracks the subject's physiological state. Specifically, the adaptive bands of respiration and heartbeat are defined in [bpm] as ${B_{\rm{adp}}^{\left( R \right)}}\left( {\hat f_R^{\left( z \right)}} \right) \triangleq \left[ \hat f_R^{\left( z \right)} - {\varepsilon _R} \hspace{0.2cm}{\hat f_R^{\left( z \right)} + {\varepsilon _R}}\right]$ and ${B_{\rm{adp}}^{\left( H \right)}}\left( {\hat f_H^{\left( z \right)}} \right) \triangleq \left[ \hat f_H^{\left( z \right)} - {\varepsilon _H} \hspace{0.2cm}{\hat f_H^{\left( z \right)} + {\varepsilon _H}}\right]$, respectively, where ${\varepsilon _R}$ and ${\varepsilon _H}$ are predefined scalars which determine the margins of the corresponding bands. Algorithm \ref{alg:E-VSDR} below summarizes the E-VSDR approach. \vspace{-0.2cm}
\begin{algorithm}[h!]
\begin{algorithmic}
\caption{Continuous NCVSM by E-VSDR}
\State \hspace{-0.45cm} \textbf{Input:}  $\{\hat{{\bf{v}}}_z\}_{z=1}^{\hat{Z}}$, $B^{\left( R \right)}$, $B^{\left( H \right)}$ 
\State \hspace{-0.3cm}\textbf{for each $z$'th detected human} $z=1,\ldots,\hat{Z}$:
\State \hspace{-0.32cm} \textbf{1:} Assemble ${\bf{D}}^{\left( R \right)}$ and ${\bf{D}}^{\left( H \right)}$ given $B^{\left( R \right)}$ and $B^{\left( H \right)}$ (\ref{D_def}), (\ref{g_q_freqs})
\State \hspace{-0.35cm} \textbf{2:} ${\cal{S}}_R^{(z)}= \mathop {\arg \max }\limits_{q} \{| ({{\bf{D}}^{\left( R \right)T}}{\hat{{\bf{v}}}_z})_q| \}:\hat{f}_R^{\left( z \right)}=g_q^{\left( R \right)}|_{q={\cal{S}}_R^{(z)}}$
\State \hspace{-0.35cm} \textbf{3:} ${\bf{d}}_{{\cal{S}}_R}^{(z)}={\bf{D}}^{\left( R \right)}(:,{{\cal{S}}_R^{(z)}})$ and $\hat{a}_{{\cal{S}}_R}^{(z)}={\bf{d}}_{{\cal{S}}_R}^{(z)T}{\hat{{\bf{v}}}_z}/({\bf{d}}_{{\cal{S}}_R}^{(z)T}{\bf{d}}_{{\cal{S}}_R}^{(z)})$
\State \hspace{-0.35cm} \textbf{4:} $    {{\hat{{\bf{v}}}}'_z} = {\hat{{\bf{v}}}_z}-{\bf{d}}_{{\cal{S}}_R}^{(z)}\hat{a}_{{\cal{S}}_R}^{(z)}$ 
\State \hspace{-0.35cm} \textbf{5:} Assemble ${{\bf{D}}^{\left( {R',z} \right)}}$ and ${{\bf{D}}^{\left( {H',z} \right)}}$ given $\hat{f}_R^{\left( z \right)}$ (\ref{RR_hamonics_def}), (\ref{S_R}) 
\State \hspace{-0.35cm} \textbf{6:} $\hat{{\bf{a}}}_z^{({R'} )}=\left({{\bf{D}}^{\left( {R',z} \right)T}}{{\bf{D}}^{\left( {R',z} \right)}}\right)^{-1}{{\bf{D}}^{\left( {R',z} \right)T}}{{\hat{{\bf{v}}}}'_z}$  
\State \hspace{-0.35cm} \textbf{7:} $    {{\hat{{\bf{v}}}}''_z}={{\hat{{\bf{v}}}}'_z}-{{\bf{D}}^{\left( {R',z} \right)}}\hat{{\bf{a}}}_z^{({R'} )}$
\State \hspace{-0.39cm} \textbf{8:} ${\cal{S}}_H^{(z)}= \mathop {\arg \max }\limits_{q} \{ |({{{\bf{{D}}}^{\left( {H',z} \right)T}}{\hat{{\bf{v}}}''_z}})_q| \}:\hat{f}_H^{\left( z \right)}=g_q^{\left( H',z \right)}|_{q={\cal{S}}_H^{(z)}}$
\State \hspace{-0.38cm} \textbf{9:} If current monitoring time $ t\geq T_{\textrm{ref}}$, than perform Signal Refinement (Subsection \ref{signal_refinement}) and replace the bands:\\ $B^{\left( R \right)}={B_{\rm{adp}}^{\left( R \right)}}\left( {\hat f_R^{\left( z \right)}} \right)$ and $B^{\left( H \right)}={B_{\rm{adp}}^{\left( H \right)}}\left( {\hat f_H^{\left( z \right)}} \right) $
\State \hspace{-0.3cm}{\bf{end for}}

\State \hspace{-0.4cm}\textbf{Output:} Refined $\{{\hat f_R^{\left( z \right)}},{\hat f_H^{\left( z \right)}}\}_{z=1}^Z$
\label{alg:E-VSDR}
\end{algorithmic}
\end{algorithm}

\newpage
Algorithm \ref{alg_robust_multi_person_NCVSM} below outlines our complete approach for robust multi-person localization and vital signs monitoring in low-SNR, cluttered environments using MIMO FMCW radar, based on the model described in Section \ref{sec:Model}.
\begin{algorithm}[h!]
\begin{algorithmic}
\caption{Robust Multi-Person Localization and Vital Signs Monitoring Using MIMO FMCW Radar}
\State \hspace{-0.4cm} \textbf{Input:} $T_{\textrm{loc}}$, $T_{\textrm{int}}$, ${T_{\textrm{win}}}$, $\{y\left[ {n,k,l} \right] \}$, $\gamma$, ${L_f}$, $I_{\textrm{max}}$, $B^{\left( R \right)}$, $B^{\left( H \right)}$
\State \hspace{-0.3cm}\textbf{At first} $T_{\textrm{loc}}$ \textbf{do:}
\State \hspace{-0.02cm} \textbf{1:} Assemble $\{{\bf{Y}}_l\}_{l=1}^{L={T_{\textrm{loc}}}f_s}$, $\bf{A}$ and $\bf{B}$ (\ref{Y=AXB+W})
\State \hspace{0.0cm} \textbf{2:} Filter $\{{\bf{Y}}_l\}_{l=1}^L$ by (\ref{Y_k_filtered}) and recover $\{{\bf{X}}_l\}_{l=1}^L$ and $\mathcal{S}$ using\\       \hspace{0.35cm} RaLU-JSR (Algorithm \ref{alg_loc_3D_JSR_FISTA})
\State \hspace{-0.4cm} \textbf{Output:} $\mathcal{S}$ $=>$ $\hat{Z}$ and $\{\hat{d}^{(z)},\hat{\theta}^{(z)}\}_{z=1}^{\hat{Z}}$ 
\State \hspace{-0.4cm} \textbf{After} $T_{\textrm{win}}$, \textbf{for each} $T_{\textrm{int}}$ \textbf{do:}
\State \hspace{0.0cm} \textbf{1:} Assemble $\{{\bf{Y}}_l\}_{l=1}^{L={T_{\textrm{win}}}f_s}$  (\ref{Y=AXB+W})
\State \hspace{0.0cm} \textbf{3:} Use $\mathcal{S}$ to evaluate $\{{\hat x_{S\left( z \right)}}\left[ l \right]\}_{z=1}^{\hat{Z}} $ (\ref{x_s_z}) and $\{\hat{{\bf{v}}}_z\}_{z=1}^{\hat{Z}} $ (\ref{v_z_est})
\State \hspace{-0.0cm} \textbf{4:} Estimate $\{ {f_H^{\left( z \right)},f_R^{\left( z \right)}}\}_{z = 1}^Z$ given $\{\hat{{\bf{v}}}_z\}_{z = 1}^{\hat{Z}}$, $B^{\left( R \right)}$ and $B^{\left( H \right)}$ \\
   \hspace{0.35cm} using E-VSDR (Algorithm \ref{alg:E-VSDR}))
\State \hspace{-0.4cm} \textbf{Output:}  $\{ {\hat{f}_H^{\left( z \right)},\hat{f}_R^{\left( z \right)}}\}_{z = 1}^{\hat{Z}}$
\label{alg_robust_multi_person_NCVSM}
\end{algorithmic}
\end{algorithm}

\section{Custom Multi-Person Hardware Phantom}
\label{sec:phantom}
\begin{figure}[htbp!]  
\begin{center}
\subfigure{\includegraphics[width=0.35\textwidth]{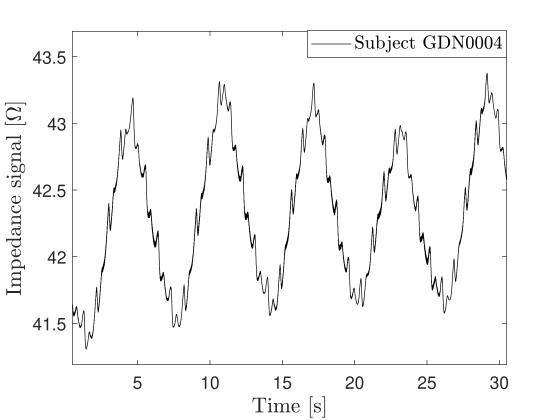}}\hspace{-0.4cm}
\end{center} 
\vspace{-0.2cm}
\caption{Example of impedance signal from \cite{schellenberger2020dataset} that was used to generate realistic mechanical displacements for the vibration unit corresponding to changes in human thoracic volume.}   
\label{fig:Impedance}
\vspace{-0.4cm}
\end{figure}

As outlined in the Introduction, we designed a custom hardware phantom capable of simulating multi-person NCVSM in real-world, cluttered environments. The phantom utilizes recorded impedance signals, that capture variations in electrical impedance across the thorax caused by changes in thoracic volume \cite{ernst1999impedance,sherwood1990methodological}. An example from a subject in \cite{schellenberger2020dataset} is shown in Fig. \ref{fig:Impedance}. These signals were selected to replicate the mechanical movements of the monitored individuals’ thoraces via a dedicated vibration unit. The phantom is designed to support a wide range of cardio-respiratory patterns, including resting, rapid, and slow breathing or heartbeat, as well as pathological conditions such as apneas, arrhythmias, and Cheyne-Stokes respirations \cite{naughton1998pathophysiology}. By replicating realistic monitoring scenarios, including those required by healthcare providers, the phantom can facilitate rigorous validation of radar systems and algorithms for both single and multi-person NCVSM, ensuring their robustness and readiness for deployment in healthcare and IoT applications.

\begin{figure}[htbp!]  
\begin{center}
\subfigure[]{\label{vib_unit}\includegraphics[width=0.24\textwidth]{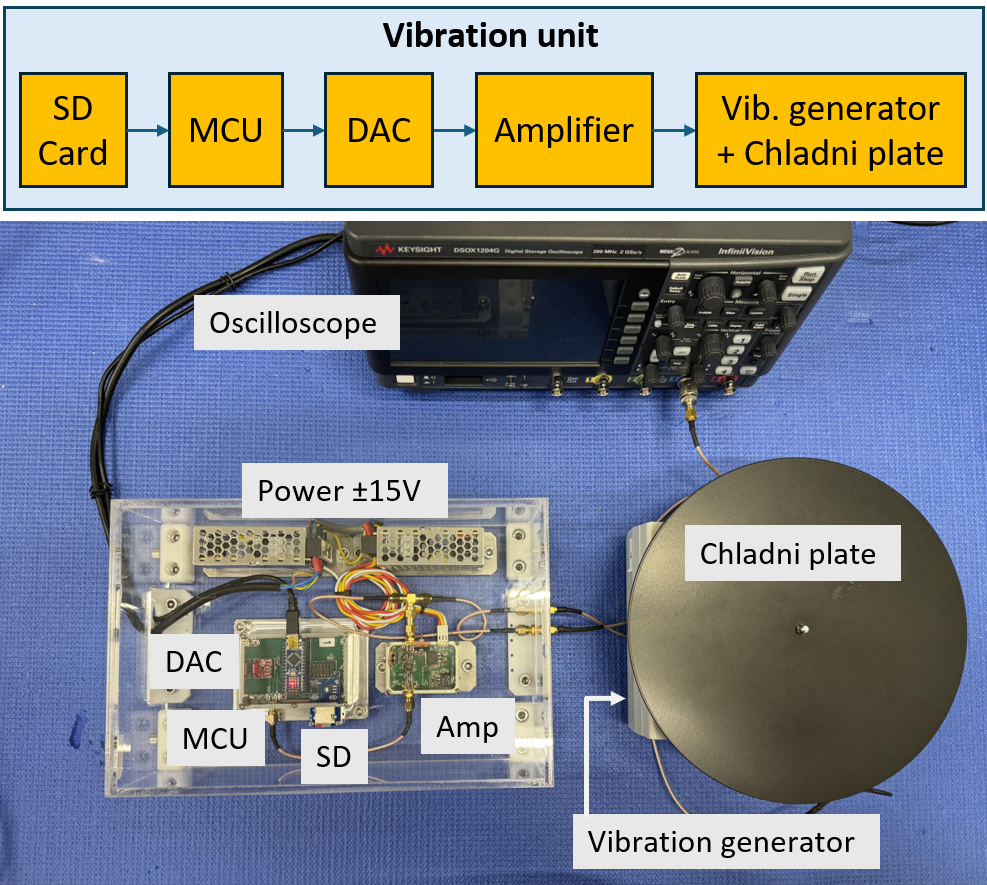}}\hspace{-0.0cm}
\subfigure[]{\label{phantom_block}\includegraphics[width=0.24\textwidth]{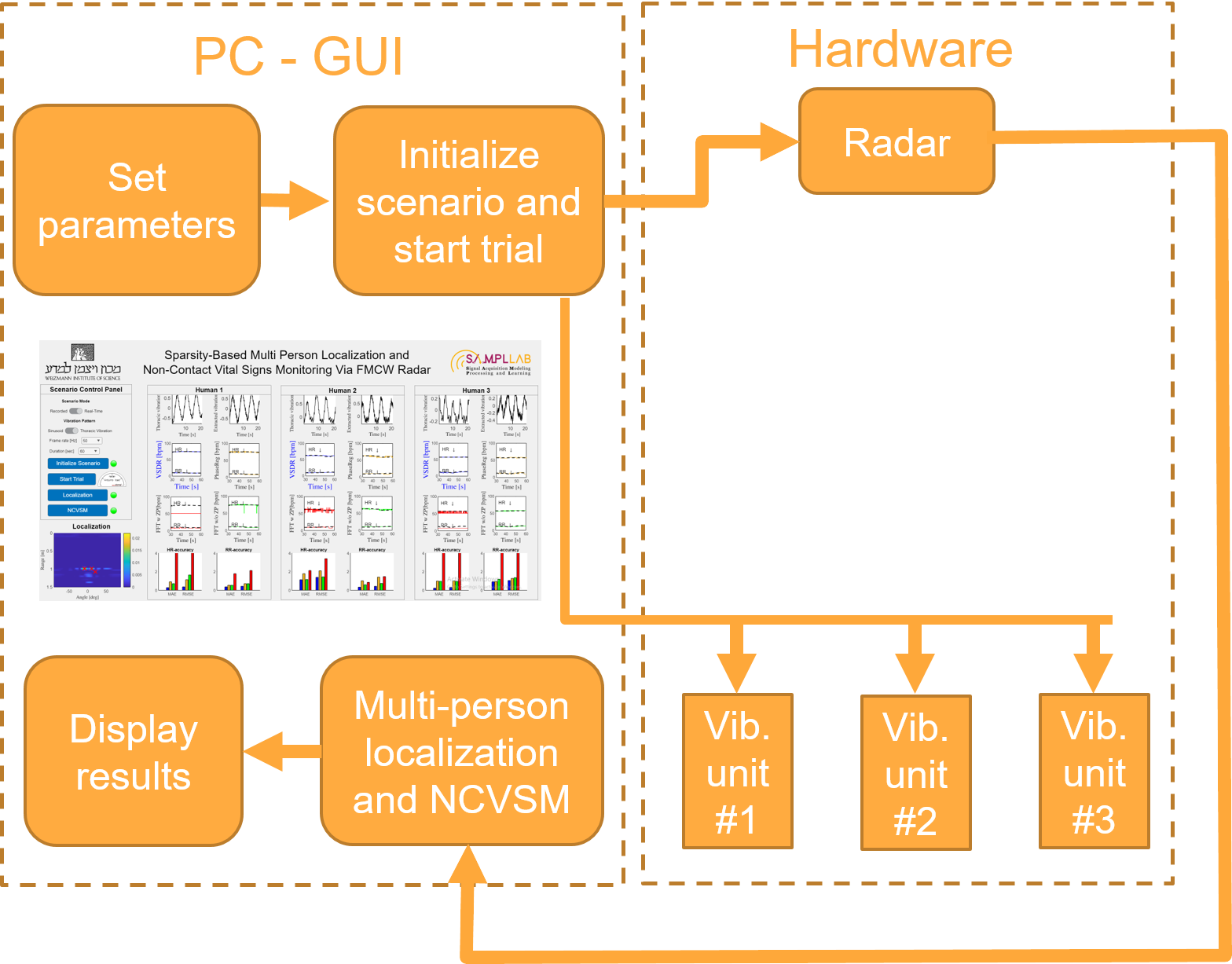}}
\end{center} 
\vspace{-0.2cm}
\caption{Block diagrams of the developed phantom. \textbf{(a)} Key components of a single vibration unit. \textbf{(b)} Overview of the validation process, comprising a radar, up to three vibration units, and a GUI for control.}   
\label{fig:phantom}
\vspace{-0.4cm}
\end{figure}

The phantom is designed in two configurations: one for single-person NCVSM, consisting of a single vibration unit (Fig. \ref{fig:exp_setup}. c$1$), and another for multi-person NCVSM, comprising three independent vibration units (Fig. \ref{fig:exp_setup}. c$3$). Each oscillating unit, as illustrated in Fig. \ref{vib_unit}, includes the following key components: $1$. Secure Digital card: Preloaded with recorded impedance signals, these cards store the digital data to be converted into mechanical oscillations.  $2$. Microcontroller unit (MCU), based on the Arduino processor \cite{arduino_nano}: Reads the digital data from the SD card and sends it to $3$. a digital-to-analog converter (DAC). $4$. High-speed operational amplifier: A specialized amplifier adjusts the analog signal's current to effectively drive $5$. a vibration generator  \cite{vibration_generator_3bscientific} fitted with a circular Chladni plate \cite{chladni_plate_3bscientific} on top.

Fig. \ref{phantom_block} depicts a block diagram of the validation process using the dedicated phantom in the multi-person setup. A graphical user interface (GUI) was developed under MATLAB environment to conveniently handle the validation stage using the hardware. This setup enables the validation of radar systems and algorithms across a wide range of scenarios, including the positioning of multiple individuals at varying distances and angles relative to the radar, and exhibiting different cardiopulmonary conditions.


\vspace{-0.3cm}

\section{Experiments and Results}
\label{sec:Performance}
This section evaluates the performance of the suggested approach and compares it to existing techniques for both single-person and multi-person NCVSM in a cluttered demonstration room, initially via the proposed custom phantom, then by human trials as elaborated below.

\vspace{-0.2cm}
\subsection{Experimental Setup}
\begin{figure}[htbp!]  
\begin{center}
\hspace{-0.3cm}
\subfigure{\label{}\includegraphics[width=0.49\textwidth]{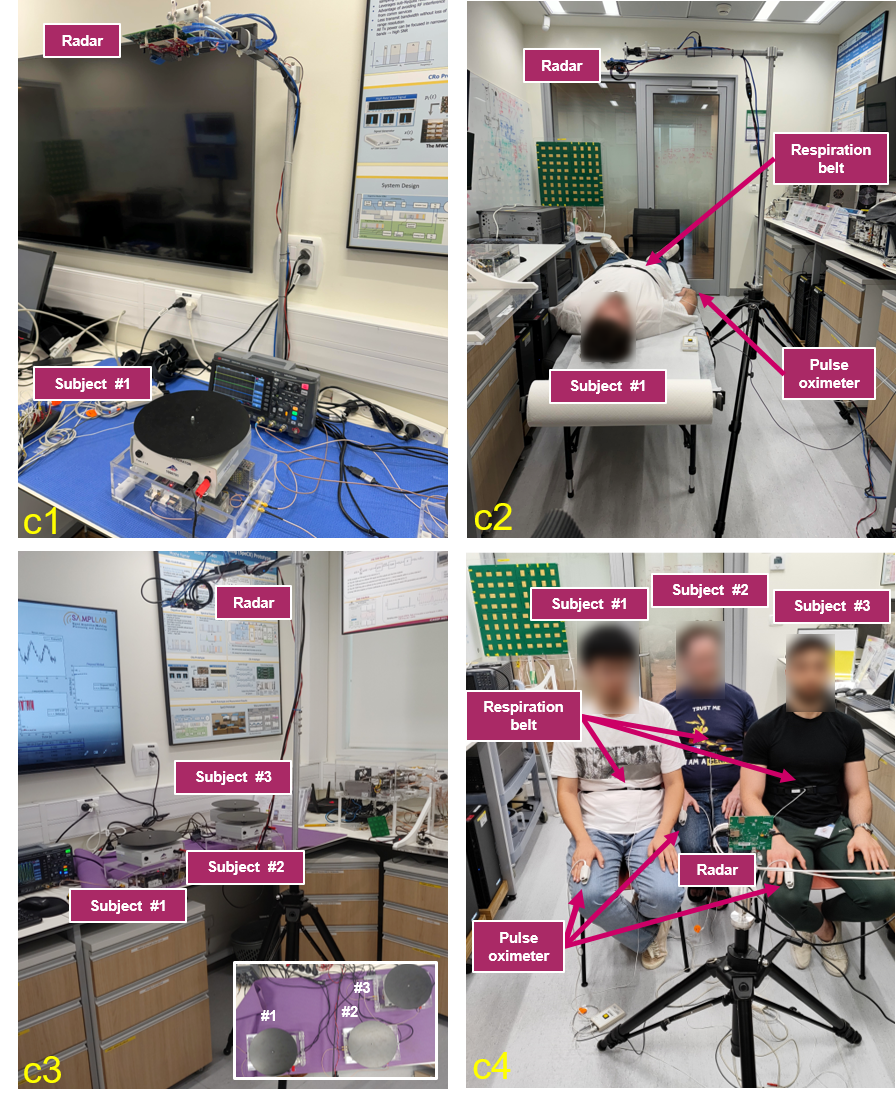}}
\end{center} 
\caption{Experimental setups in a cluttered demonstration room.\\ {(c$1$)} Single-person phantom trials. {(c$2$)} Single-person human trials. {(c$3$)} Multi-person phantom trials. {(c$4$)} Multi-person human trials.  }
\label{fig:exp_setup}
\end{figure}

The proposed approach was validated by conducting $4$ classes of experiments in a cluttered demonstration room containing multiple objects in the radar's FOV, such as computers, tables, and various electrical devices, as shown in Fig. \ref{fig:exp_setup}. The classes were set as follows: c$1$ - single-person phantom trials, c$2$ - single-person human trials (lying back on a bed), c$3$ - multi-person phantom trials, and c$4$ - multi-person human trials (sitting on chairs). In each single-person setup (c$1$/c$2$),  $9$ simulated/human subjects were examined in $9$ separate trials. In each multi-person setup (c$3$/c$4$),  $9$ simulated/human subjects were divided into $3$ experiments of $3$ people each. In every class, the targets were monitored continuously for $2$ minutes with RR and HR estimates computed every $T_{\textrm{int}}=0.05$ [s], using $L$ frames of $\{{\bf{Y}}_l\}_{l=1}^L$ (\ref{Y=AXB+W}) collected from the last $T_{\textrm{win}}=30$ [s], starting at $T_{\textrm{win}}$. The human trials were approved by the Weizmann Institutional Review Board - protocol $\#2214-1$, and informed consent was obtained from $16$ participants ($9$ males and $7$ females, height from $156$ to $183$ [cm], weight from $47$ to $87$ [kg] and age from $19$ to $54$ years old). 

In the phantom trials (c$1$ and c$3$), the simulated thoracic displacements were based on the impedance signal of $9$ individuals from the resting scenario of \cite{schellenberger2020dataset}, as exemplified in Fig. \ref{fig:Impedance}, that were converted into proper mechanical vibrations. The corresponding raw impedance signal and lead-$2$ ECG signal from \cite{schellenberger2020dataset}, respectively served as the ground-truth (GT) references for comparing the RR and HR estimates \cite{ernst1999impedance, sherwood1990methodological,turppa2020vital}, after down-sampling them to reach $T_s$. In the human trials (c$2$ and c$4$), the radar was aimed at the thorax of the subjects who were asked to breathe calmly and avoid large movements. As for the GT references of the human trials, we used the g.Hiamp device \cite{gtec_biosignal_amplifier} (FDA-cleared and CE-certified medical product) for comparing the radar-based RR and HR estimates. Specifically, the GT-RR estimates were calculated from the torso circumference signal obtained from the Respiration Effort Sensor (respiration belt) whereas the GT-HR estimates were calculated from the photoplethysmogram (PPG) signal obtained from the g.SpO2sensor (pulse oximeter) \cite{gtec_body_sensors}.

The radar system selected for this work was Texas Instruments (TI) IWR1443BOOST $76$ to $81$ [GHz] mmWave Sensor Evaluation Module (EVM) \cite{ti_iwr1443boost}, connected to DCA1000EVM \cite{ti_dca1000evm} for data capture and streaming. We employed the horizontal MIMO ULA setup of $2$ transmitting antennas and $4$ receiving antennas. In the single-person trials (c$1$ and c$2$), we improved the SNR by simultaneously using both transmitters to create a $1\times{K}$ SIMO array with increased transmission power. In the general case of monitoring multiple targets (c$3$ and c$4$), we improved the angular resolution using a TDM transmission scheme that produced a virtual antenna array of size $1\times{\tilde{K}}$ with $\tilde{K}=2K$, as depicted in Fig. \ref{fig:MIMO_TDM} for $K=4$. The main radar parameters, reflected in the signal model in (\ref{Y=AXB+W}) are summarized in Table \ref{table:FMCW_parameters}. We note that for bandwidth $B=ST_c\approx 4$ [GHz], the range resolution is $d_{\textrm{res}}=\frac{c}{2B}\approx 3.75$ [cm]. In addition, the radar parameters listed in Table \ref{table:FMCW_parameters} through (\ref{fm_dm}), (\ref{fm_Nyquist_grid}) and (\ref{theta_p_grid}) with angle grid spacing set as $\Delta_{\theta}=1$, enable coverage over a radial distance range of $d_{\rm{min}}=4.29$ [cm] to $d_{\rm{max}}=4.24$ [m] and an angular range of $\theta_{\rm{min}}=-90^\circ$ to $\theta_{\rm{max}}=+89^\circ$. 

Multi-person NCVSM often necessitates both range and angular separation of targets. To evaluate these aspects, in our multi-person trials (c$3$ and c$4$), two of the three subjects were positioned at similar radial distances from the radar but at distinct azimuth angles, while the third subject was placed at a different radial distance and azimuth angle. The exact positionings for each class are provided in Table \ref{table:MP_setup}. For better readability, the outcomes of the multi-person trials (c$3$ and c$4$) are reported herein, while those of the single-person trials (c$1$ and c$2$) are provided in the supplementary material document. 


\begin{table}[h!]
\caption{MIMO FMCW radar parameters}
\label{table:FMCW_parameters}
\centering
\begin{tabular}{ |l |c | c|  }
\hline
Parameter & Symbol & Value\\
\hline
Maximal chirp wavelength & $\lambda_{\textrm{max}}$   & $3.9$ [mm]\\
Chirp duration & $T_c$   & $57$ [$\mu \textrm{s}$]\\
ADC sampling rate & $f_{\textrm{ADC}}$   & $4$ [MHz]\\
Rate of frequency sweep & $S$   & $70$ [MHz/$\mu$s]\\
Frame duration & $T_s$   & $50$ [ms]\\
$\#$ of selected \textit{fast-time} samples  & $\bar{N}$   & $200$ \\
$\#$ of chirps per frame & $G$   & $40$ \\
$\#$ of transmitters & $J$   & $2$ \\
$\#$ of receivers (virtual) & $K$ ($\tilde{K}$)   & $4$ ($8$) \\
 \hline
\end{tabular}
\end{table}

\begin{table}[h!]
\caption{Subject positionings for each class}
\label{table:MP_setup}
\centering
\begin{tabular}{ |c |c | c| c| }
\hline
Class & Trial's subject $\#$ & Distance [m] & Angle [$^{\circ}$] \\
\hline
c$1$ & $1$ & $0.7$ & $0$ \\
\hline
c$2$ & $1$ & $1.3$ & $0$ \\
\hline
     & $1$  & $0.80$ & $-10$ \\
c$3$ & $2$  & $0.80$ & $+10$ \\
     & $3$  & $0.85$ & $-15$ \\
\hline
     & $1$  & $1.30$ & $-30$ \\
c$4$ & $2$  & $1.30$ & $+30$ \\
     & $3$  & $1.80$ & $0$ \\
\hline
\end{tabular}

\end{table}

\vspace{-0.3cm}
\subsection{Localization Setup}
As discussed in the Introduction, it is necessary to achieve precise localization of all persons' thorax to accurately monitor their vital signs from the extracted thoracic vibrations. For every trial, we assessed the localization map given data from only the first five seconds of the monitoring session. That is, given $\{{\bf{Y}}_l\}_{l=1}^L$ (\ref{Y=AXB+W}) assembled by the first $L=f_sT_{{\rm{loc}}}$ frames with $T_{{\rm{loc}}}=5$ [s]. Here, we compared the estimated $2$D range-angle map ${\bf{\bar{X}}}$ by the proposed RaLU-JSR (Algorithm \ref{alg_loc_3D_JSR_FISTA}) to that obtained using the \textit{Angle-FFT} method used in \cite{ahmad2018vital,gao2019experiments,han2021detection}, the \textit{MUSIC} approach \cite{konno2014experimental,kim2020low} employed for each FFT range-bin, the second-order differential extension called \textit{SOD-MUSIC} \cite{xiong2022vital}, the linear constrained minimum variance-based adaptive beamforming, here called \textit{LCMV} \cite{wang2021multi}, and the calibrated CIR technique \cite{wang2021driver} with $5$ past CIR maps to average in each frame, here called \textit{cal-CIR}. 

The parameters of RaLU-JSR were set as follows: The number of iterations and regularization parameter were selected to $I=1000$ and $\gamma=100$, respectively. Following the Lipschitz relation below Algorithm \ref{alg_loc_3D_JSR_FISTA}, the discretized matrices ${\bf{A}}$ and ${\bf{B}}$ (Subsection \ref{Pre_Proc}) lead to $L_f =  8.8439e+04$ for the multi-person analysis when $\tilde{K}=8$ and $L_f =  6.3750e+04$ for the single-person analysis when $K=4$. Finally, the vital frequencies of $\bf{\Pi}$ in (\ref{Y_k_filtered}) were drawn from the length-$L$ \textit{slow-time} Nyquist grid determined by $f_s$ according to the respiration and heartbeat bands ${B^{\left( R \right)}}=\left[0.1\hspace{0.2cm}0.5 \right]$ [Hz] and
${B^{\left( H \right)}}=\left[0.83\hspace{0.2cm}1.67 \right]$ [Hz], respectively, corresponding to a normal resting state.

For a fair comparison, all maps were produced using identical range and angle grids, were normalized by the respective maximum value within a designated ROI of [$0.5$ $2$] [m] and [$-50$ $+50$] [$^{\circ}$] and were slightly denoised by setting values below $0.05\%$ of the maximum to zero. Here, we determined $\mathcal{S}$ by $2$D selection of peaks that exceed a normalized power threshold of $0.1$ and $0.4$ for range and angle, respectively. In each localization figure, cyan circles (O) indicate the true thoracic locations, while red crosses (X) represent the estimated positions based on the proposed detection scheme.

\subsection{NCVSM Setup}
After the individuals are accurately located, their vital signs are monitored continuously for every $T_{\rm{int}}$, given $\cal{S}$ and the last $L=f_sT_{\rm{win}}$ frames of $\{{\bf{Y}}_l\}_{l=1}^L$ (\ref{Y=AXB+W}) collected up to that time.

We evaluated the performance of the E-VSDR method (Algorithm \ref{alg:E-VSDR}) for both single-person and multi-person NCVSM given $\{\hat{{\bf{{v}}}}_z\}_{z=1}^{\hat{Z}}$ (\ref{v_z_est}) and the vital bands ${B^{\left( R \right)}}$ and ${B^{\left( H \right)}}$. Recall that the last stage of the E-VSDR incorporates a signal refinement procedure designed to enhance accuracy in continuous NCVSM, by mitigating noise and possible overlaps between HR and RR harmonics. The parameters of the refinement were set to $T_{\rm{ref}}=5$ [s], $T_{\rm{avg}}^{(H)}=3$ [s], $T_{\rm{avg}}^{(R)}=5$ [s] and $\varepsilon_H=\varepsilon_R=5$. The results of the E-VSDR were compared to those obtained using several state-of-the-art NCVSM techniques: $1$. The method detailed in \cite{adib2015smart} for estimating RR and HR given the phase of a FMCW signal, called here \textit{PhaseReg}. $2$. FFT-based peak selection in each frequency band \cite{sacco2020fmcw,alizadeh2019remote,antolinos2020cardiopulmonary}, termed here \textit{FFT}. $3$. The approach suggested in \cite{xiong2022vital} which employs a FFT-based peak selection on the extracted phase after the removal of high-order respiration harmonics via orthogonal projections, named here \textit{OrthProj}. To assess the impact of the refinement on the results, the performance was also compared with these three techniques augmented with the same refinement procedure used in E-VSDR. We refer them to as $4$. \textit{PhaseReg+}, $5$. \textit{FFT+} and $6$. \textit{OrthProj+}, respectively. As for the reference data, the GT-RR and GT-HR were calculated from the raw data of the contact sensors via the DFT spectrum \cite{adib2015smart,turppa2020vital,mercuri2019vital,sacco2020fmcw,alizadeh2019remote}, here padded to fit a $60$-second time window to correspond to an optimal frequency resolution of $1$ [bpm].

To compare the NCVSM methods fairly, we assume that all considered subjects were accurately detected and positioned. That is, the following comparison was performed given the same extracted thoracic vibrations $\{\hat{{\bf{v}}}_z\}_{z=1}^{\hat{Z}}$ (\ref{v_z_est}) using the true locations of the examined subjects. In addition, all methods used the same vital frequency bands ${B^{\left( R \right)}}$ and ${B^{\left( H \right)}}$, and all other settings that preceded this step were the same. To assess the accuracy of the estimation methods w.r.t. the references, we used the following evaluation metrics for each class: $1$. Average empirical Cumulative Distribution Function (AeCDF), defined here as the average percentage of instances in which the estimate deviated from the reference output during a monitoring session (Y-axis) by less than a variable error threshold of bpm (X-axis), over all trials within the class. For example, given an error threshold of $2$ [bpm], the corresponding value on the Y-axis represents the average proportion of the monitoring period in which the estimations fell within +-$2$ [bpm] relative to the references, also called success rate - $2$ [bpm] \cite{eder2023sparsity}. $2$. Root-Mean-Square Error (RMSE) analysis for each class, including subject-specific values along with the average and median of the specified class.

\subsection{Multi-Person Phantom Trials}
Since the phantom validations served as an intermediate step to determine the optimal parameters and configurations prior to conducting human trials, we begin by analyzing the results from the phantom trials (here multi-person c$3$). The analysis is conducted in three stages: localization, followed by NCVSM, and concluding with a performance evaluation. 

Fig. \ref{fig:MP_loc_h} below depicts the compared localization maps of trial $\#2$ from c$3$, normalized within the designated ROI. We analyze the results of each method from left to right of each row: $1$. Although the \textit{Angle-FFT} detected some reflections from the subjects, these were smeared due to resolution limitations and mispositioned, as the method was biased toward stronger reflections originating from the table supporting the plates. $2$. The \textit{MUSIC} approach seeks highly reflecting or oscillating objects in each range-bin. Hence, in our cluttered scenario, it mistakenly highlighted clutter over humans, which deteriorated performance in both detection and positioning. $3$. The \textit{SOD-MUSIC} map was considerably cleaner than its predecessor. However, the influence of clutter remained substantial, hindering the correct localization of the subjects. $4$. For the \textit{LCMV} map, the table completely masked the presence of the targets. $5$. The \textit{cal-CIR} map resembled that of \textit{Angle-FFT} in its ability to identify certain reflections from the phantom. However, these reflections were far weaker than those from the table, leading to inaccurate localization. $6$. In contrast to the compared localization techniques, the proposed RaLU-JSR detected the right number of simulated subjects and their location, without range error and with angular error of less than $3$ [$^{\circ}$], far below the theoretical resolution limitation of $\approx15$ [$^{\circ}$] when using $8$ receivers \cite{gao2019experiments}. These results stem from the fact that RaLU-JSR uniquely leverages both human vital frequencies and the sparse properties of the data through the proposed bilinear model in (\ref{Y=AXB+W}). 

\begin{figure*}[htbp!]  
\begin{center}
\subfigure{\includegraphics[width=1.00\textwidth]{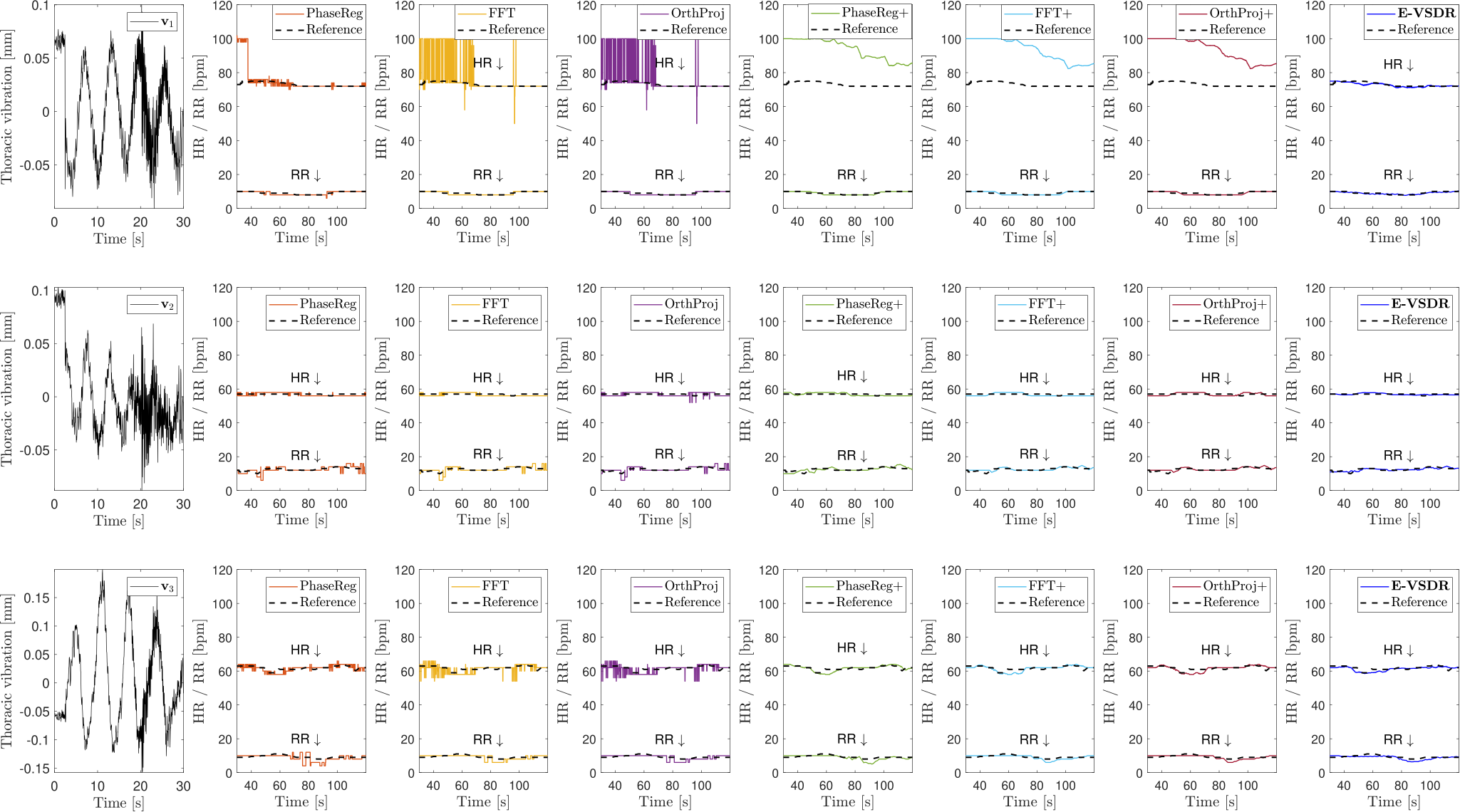}}\vspace{-0.2cm}
\end{center} 
\vspace{-0.2cm}
\caption{NCVSM of $3$ subjects for multi-person phantom trial - class c$3$, trial $\#2$. \textbf{Rows:} Subjects $1$-$3$. \textbf{Columns:} Extracted thoracic vibrations ${\bf{v}}_1$-${\bf{v}}_3$ for some $T_{\textrm{int}}$, \textit{PhaseReg} \cite{adib2015smart}, \textit{FFT} \cite{sacco2020fmcw,alizadeh2019remote,antolinos2020cardiopulmonary}, \textit{OrthProj} \cite{xiong2022vital} and the refined versions \textit{PhaseReg+}, \textit{FFT+}, and \textit{OrthProj+}. The rightmost plots show the proposed E-VSDR estimates, which demonstrate the closest alignment with the reference curves compared to the competing approaches, even when aided by the proposed refinement procedure.}   
\label{fig:MP_NCVSM_h}
\vspace{-0.4cm}
\end{figure*}

Fig. \ref{fig:MP_NCVSM_h} depicts the NCVSM outcomes of this trial. The results were produced by all $7$ compared methods relative to GT references, given the corresponding vibrations $\{\hat{{\bf{{v}}}}_z\}_{z=1}^{\hat{Z}=3}$. We observed several key findings. First, since the resting heartbeat typically spans a relatively wide frequency band (here, $0.83$ to $1.67$ [Hz]), this can lead to interference from RR harmonics competing with the true HR during the frequency search. This is evident in the case of $\hat{\bf{v}}_1$ (first row), where both the original compared methods (\textit{PhaseReg}, \textit{FFT}, and \textit{OrthProj}) and their refined versions (\textit{PhaseReg+}, \textit{FFT+}, and \textit{OrthProj+}) were significantly affected by the large variability in HR estimations. The considerable fluctuation is particularly detrimental to the refined variants, as they constrict the frequency search area based on the median outcome of the onset measurements, which deviated significantly from the true values. Additionally, for $\hat{\bf{v}}_2$ and $\hat{\bf{v}}_3$ (the second and third rows), the estimation curves of the refined methods exhibited patterns with reduced noise and more closely matched the reference curves compared to their original versions. However, only the HR and RR estimates by the proposed E-VSDR demonstrated a high degree of similarity to the reference values across all three subjects.

\begin{figure*}[htbp!]  
\begin{center}
\subfigure[]{\label{fig:performance_AeCDF_h}\includegraphics[width=0.25\textwidth]{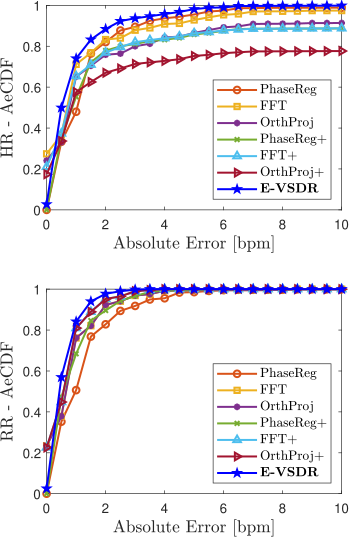}} 
\subfigure[]{\label{fig:performance_RMSE_h}\includegraphics[width=0.74\textwidth]{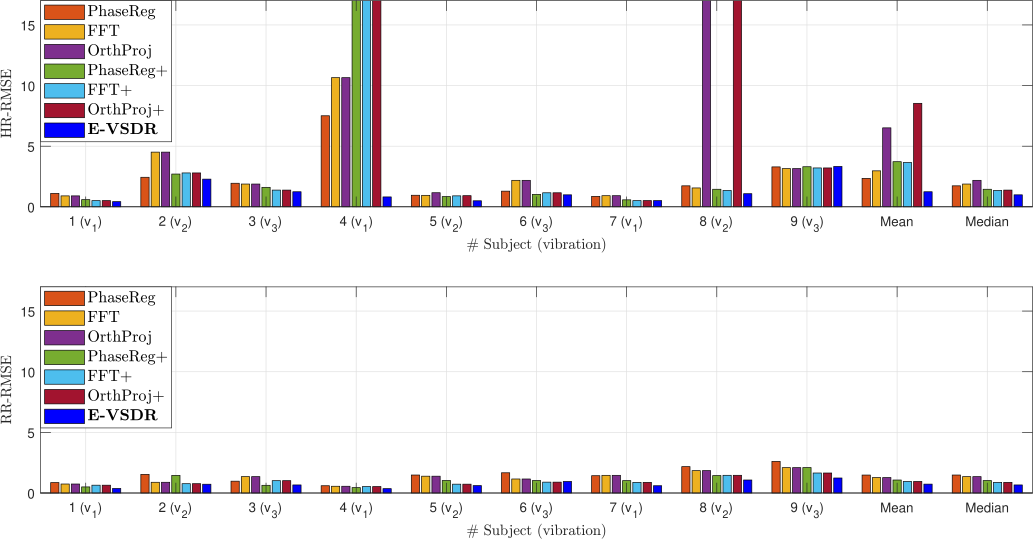}}
\end{center} 
\vspace{-0.2cm}
\caption{NCVSM performance plots for multi-person phantom trials c$3$. \textbf{(a)} Average empirical CDFs for HR and RR estimations. \textbf{(b)} RMSE scores: subject-wise with class average and median. For both HR and RR estimations, the proposed E-VSDR outperformed the compared methods. In terms of AeCDF, it surpassed for every error above $0.5$ [bpm], and in terms of RMSE, it achieved the lowest average and median values.}   
\label{fig:performance_h}
\vspace{-0.5cm}
\end{figure*}

Fig. \ref{fig:performance_AeCDF_h} shows the $9$ subjects-AeCDF of class c$3$. The AeCDF was calculated for both HR and RR estimations by all examined methods, as a function of absolute error thresholds ranging from $0$ to $10$ [bpm]. First, one sees that the E-VSDR outperforms all other compared methods for every error threshold greater than $0.5$ [bpm]. For instance, an average success rate of $2$ (ASR2), $3$ (ASR3) and $4$ (ASR4) [bpm] was exclusively reached by our E-VSDR method for accuracy of $88.41\%$, $93.66\%$ and $95.74\%$, respectively, for HR estimation and accuracy of $97.74\%$, $99.77\%$ and $100\%$, respectively, for RR estimation. For the numerical values obtained by all compared methods, see Table \ref{table:accuracy}. Interestingly, the refinements applied to the competing methods primarily improved the RR performance, although they still did not surpass the E-VSDR. While the E-VSDR demonstrated the best overall performance in the empirical CDF analysis, the relatively small performance gap, particularly in RR estimations, can be attributed to the phantom mechanism's reliance on impedance signals with prominent cardiopulmonary information, which facilitates adequate estimation performance across all methods.

Finally, Fig. \ref{fig:performance_RMSE_h} presents the HR-RMSE and RR-RMSE distributions for each NCVSM method across the 9 subjects of class c$3$, along with the corresponding average and median values. The proposed E-VSDR achieved the lowest average and median RMSE values for both HR and RR estimations, even when using the outlier-tolerant median metric. Specifically, the class average RMSE (ARMSE) was as low as $1.23$ and $0.73$ for HR and RR estimation, respectively, with the values of the compared techniques shown in Table \ref{table:accuracy}. Additionally, E-VSDR obtained superior RMSE scores for most subjects, regardless of their location relative to the radar, with only minor deviations observed for the remaining subjects. In terms of median RMSE, the refinements applied to \textit{PhaseReg}, \textit{FFT}, and \textit{OrthProj} enhanced performance for both HR and RR estimations; however, they did not outperform the E-VSDR, which underscores the strength of the core harmonics-resilient, dictionary-based approach of the E-VSDR. Both the localization and NCVSM results via the phantom trials instilled confidence to advance to the human trials with the finalized algorithm and parameters as well as proper trial configurations.


\begin{figure*}[htbp!]  
\begin{center}
\subfigure{\includegraphics[width=0.9\textwidth]{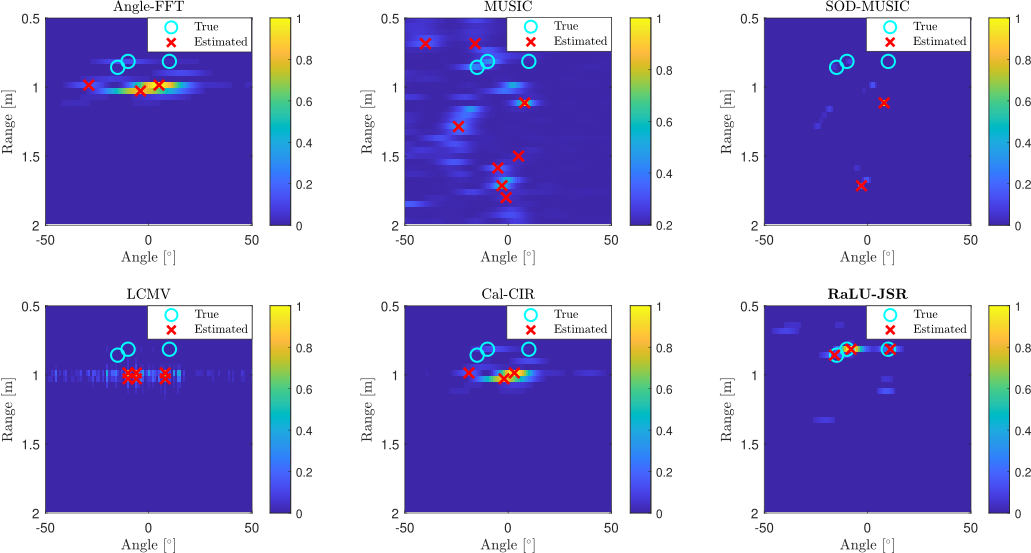}}\vspace{-0.2cm}
\end{center} 
\vspace{-0.2cm}
\caption{Localization maps of multi-person phantom trial - class c$3$, trial $\#2$. The maps were produces by \textit{Angle-FFT} \cite{ahmad2018vital,gao2019experiments,han2021detection}, \textit{MUSIC} \cite{konno2014experimental,kim2020low}, \textit{SOD-MUSIC} \cite{xiong2022vital}, \textit{LCMV} \cite{wang2021multi}, \textit{cal-CIR} \cite{wang2021driver} and the proposed RaLU-JSR (Algorithm \ref{alg_loc_3D_JSR_FISTA}). The X and O signs denote the estimated and true locations of humans, respectively. Only the proposed RaLU-JSR approach properly detects and positions all $3$ subjects in the specified scenario. }   
\label{fig:MP_loc_h}
\vspace{-0.4cm}
\end{figure*}

\begin{table}[h!]
\centering
\caption{Average success rate [$\%$] for $2$ (ASR2), $3$ (ASR3) and $4$ (ASR4) [bpm] as well as average root-mean-squared-error (ARMSE) for HR and RR estimations by the compared NCVSM methods, for multi-person classes c$3$ and c$4$.}
\begin{tabular}{ |c|c|c|c|c|c|c|}
\hline
           Class & Rate            & Method & ASR2 & ASR3 & ASR4 & ARMSE\\
\hline
                        & \multirow{6}{*}{HR}   & \textit{PhaseReg} & 82.02& 89.48 & 93.66 & 2.34\\
                        \cline{3-7}
\multirow{12}{*}{c$3$}   &  & \textit{FFT} & 83.25 & 87.83 & 90.85 & 2.96\\
                        \cline{3-7}
                        &  & \textit{OrthProj} & 75.85 &  80.00 & 84.30& 6.50\\
                        \cline{3-7}
                        &  & \textit{PhaseReg+} & 77.88 & 81.14& 83.18 & 3.71\\
                        \cline{3-7}
                        &  & \textit{FFT+} & 76.91 & 81.99 & 83.94 &  3.65\\
                        \cline{3-7}
                        &  & \textit{OrthProj+} & 67.04 & 71.33 &  72.83 & 8.53\\
                        \cline{3-7}
                        &  & \textbf{E-VSDR} & \textbf{88.41} & \textbf{93.66} & \textbf{95.74} & \textbf{1.23}\\
\hhline{|~|=|=|=|=|=|=|}
                        & \multirow{6}{*}{RR} & \textit{PhaseReg} & 82.81 & 91.80 & 95.54 & 1.47\\
                        \cline{3-7}
                        &  & \textit{FFT} & 92.27 & 96.87 & 99.20 & 1.27\\
                        \cline{3-7}
                        &  & \textit{OrthProj} & 92.27 & 96.87 & 99.20 & 1.27\\
                        \cline{3-7}
                        &  & \textit{PhaseReg+} & 89.57 & 96.65 & 99.24 & 1.06\\
                        \cline{3-7}
                        &  & \textit{FFT+} & 95.20 & 99  & 100 & 0.95\\
                        \cline{3-7}
                        &  & \textit{OrthProj+} & 95.20 & 99  & 100 & 0.95\\
                        \cline{3-7}
                        &  & \textbf{E-VSDR} & \textbf{97.74} & \textbf{99.77} & \textbf{100} & \textbf{0.73}\\   
\hhline{|=|=|=|=|=|=|=|}
                        & \multirow{6}{*}{HR} & \textit{PhaseReg} & 61.77 & 72.55 & 77.91 & 7.43\\
                        \cline{3-7}
\multirow{12}{*}{c$4$}   &  & \textit{FFT} & 57.91 & 63.30 & 66.87 & 10.03\\
                        \cline{3-7}
                        &  & \textit{OrthProj} & 57.12 & 62.48 & 66.04 & 10.11\\
                        \cline{3-7}
                        &  &\textit{PhaseReg+} & 71.58 & 75.30 & 80.23 & 4.01 \\
                        \cline{3-7}
                        &  & \textit{FFT+} & 70.01 & 79.84 & 83.26 & 5.20\\
                        \cline{3-7}
                        &  & \textit{OrthProj+} & 70.60 & 80.60 & 84.53 & 5.02\\
                        \cline{3-7}
                        &  & \textbf{E-VSDR} & \textbf{87.10} & \textbf{94.12} & \textbf{95.54} & \textbf{1.33}\\  
\hhline{|~|=|=|=|=|=|=|}
                        & \multirow{6}{*}{RR} & \textit{PhaseReg} & 68.88 & 79.51 & 88.33 & 2.46\\
                        \cline{3-7}
                        & &\textit{FFT} & 81.01 & 88.99 & 92.01 & 2.02\\
                        \cline{3-7}
                        & &\textit{OrthProj} & 81.01 & 88.99 & 92.01 & 2.02\\
                        \cline{3-7}
                        & &\textit{PhaseReg+} & 78.36 & 87.46 & 92.53 & 1.95\\
                        \cline{3-7}
                        & &\textit{FFT+} & 86.66 & 93.19 & 95.87 &  1.51\\
                        \cline{3-7}
                        & &\textit{OrthProj+} & 86.66 & 93.19 & 95.87 &  1.51\\
                        \cline{3-7}
                        & & \textbf{E-VSDR} & \textbf{94.14} & \textbf{98.12} & \textbf{98.69} & \textbf{0.98}\\   
\hline
\end{tabular}
\label{table:accuracy}
\end{table}

\subsection{Multi-Person Human Trials}
After drawing significant conclusions from the phantom trials, we proceeded with the human trials using similar parameters and configurations. The corresponding localization and NCVSM results for class c$4$ (multi-person human trials) are presented below. 

Fig. \ref{fig:MP_loc_t} below depicts the compared normalized localization maps within the designated ROI of class c$4$ - trial $\#3$. One can observe the following: $1$. While the \textit{Angle-FFT} successfully detected the presence of all three individuals, it exhibited coarse angular errors, particularly for the equidistant subjects at $1.30$ [m]. $2$. Although \textit{MUSIC} managed to locate two out of the three subjects, its high sensitivity caused clutter artifacts to emerge, which led to the miss detection of the third one. $3$. The enhanced \textit{SOD-MUSIC} algorithm improved upon \textit{MUSIC} by sharpening the output map and mitigating clutter. However, it also failed to detect one of the subjects. $4$. For the \textit{LCMV}, while the distance estimates were reasonably accurate, the method generated too many angular candidates, causing the solution to converge to the center and thus fail to resolve the equidistant individuals at $1.30$ [m]. $5$. The \textit{cal-CIR} map provided an accurate depiction of the subjects' locations in terms of both distance and angle. However, its limited ability to differentiate between clutter and humans resulted in the selection of non-human objects alongside the actual subjects. $6$. In contrast, only the suggested RaLU-JSR detected precisely all $3$ humans and identified the position of their thorax, with an angular error of less than $5$ [$^{\circ}$] for each subject, demonstrating distinguished performance in both detection and positioning. 

\begin{figure*}[htbp!] 
\vspace{-1cm}
\begin{center}

\subfigure{\includegraphics[width=1.00\textwidth]{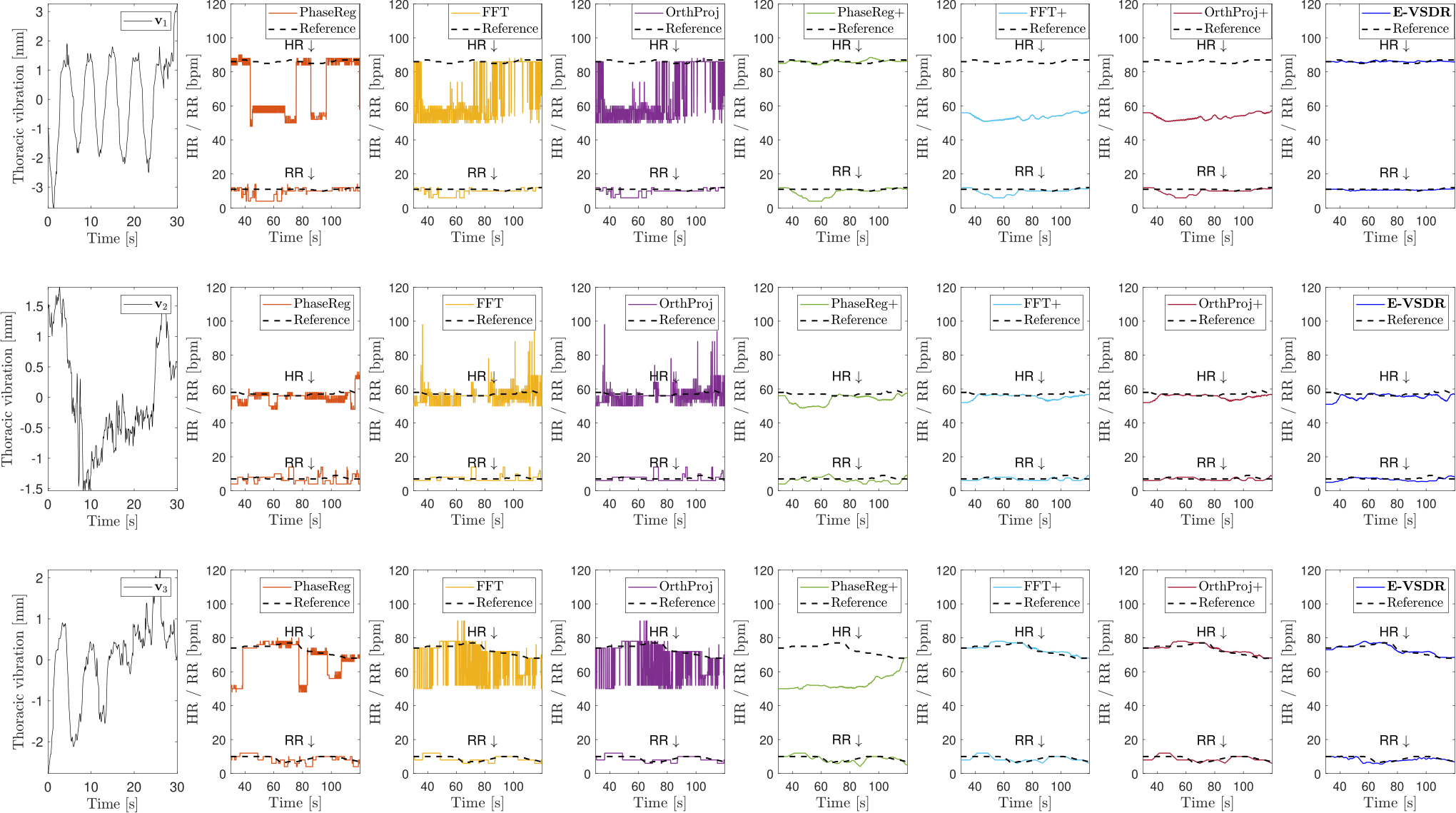}}\vspace{-0.2cm}
\end{center} 
\vspace{-0.2cm}
\caption{NCVSM of $3$ subjects for multi-person human trial - class c$4$, trial $\#3$. \textbf{Rows:} Subjects $1$-$3$. \textbf{Columns:} Extracted thoracic vibrations ${\bf{v}}_1$-${\bf{v}}_3$ for some $T_{\textrm{int}}$, \textit{PhaseReg} \cite{adib2015smart}, \textit{FFT} \cite{sacco2020fmcw,alizadeh2019remote,antolinos2020cardiopulmonary}, \textit{OrthProj} \cite{xiong2022vital} and the refined versions \textit{PhaseReg+}, \textit{FFT+}, and \textit{OrthProj+}. The rightmost plots show the proposed E-VSDR estimates, which demonstrate the closest alignment with the reference curves compared to the competing approaches, even when aided by the proposed refinement procedure.}   
\label{fig:MP_NCVSM_t}
\vspace{-0.4cm}
\end{figure*}

Fig. \ref{fig:MP_NCVSM_t} presents the NCVSM outcomes of this trial. One can notice significant variability in HR estimates among the compared methods (\textit{PhaseReg}, \textit{FFT}, and \textit{OrthProj}) across all three subjects. This variability resulted in persistent drifts in the refined versions: \textit{FFT+} and \textit{OrthProj+} for $\hat{\bf{v}}_1$ (first row) and \textit{PhaseReg+} for $\hat{\bf{v}}_2$ and $\hat{\bf{v}}_3$ (second and third rows). In contrast, the proposed E-VSDR demonstrated remarkable robustness, effectively handling the challenging noise introduced by low SNR, multipath effects, possible RBMs, clutter, and interfering harmonics, for all three subjects.

\begin{figure*}[htbp!]  
\begin{center}
\subfigure[]{\label{fig:performance_AeCDF_t}\includegraphics[width=0.25\textwidth]{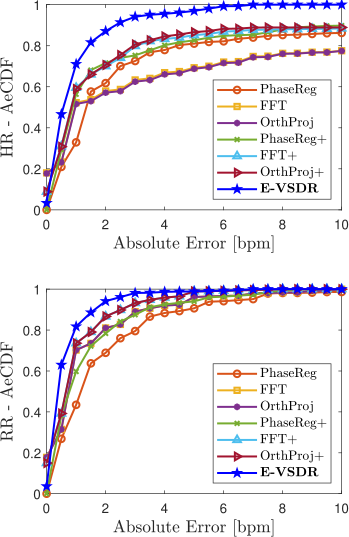}}
\subfigure[]{\label{fig:performance_RMSE_t}\includegraphics[width=0.74\textwidth]{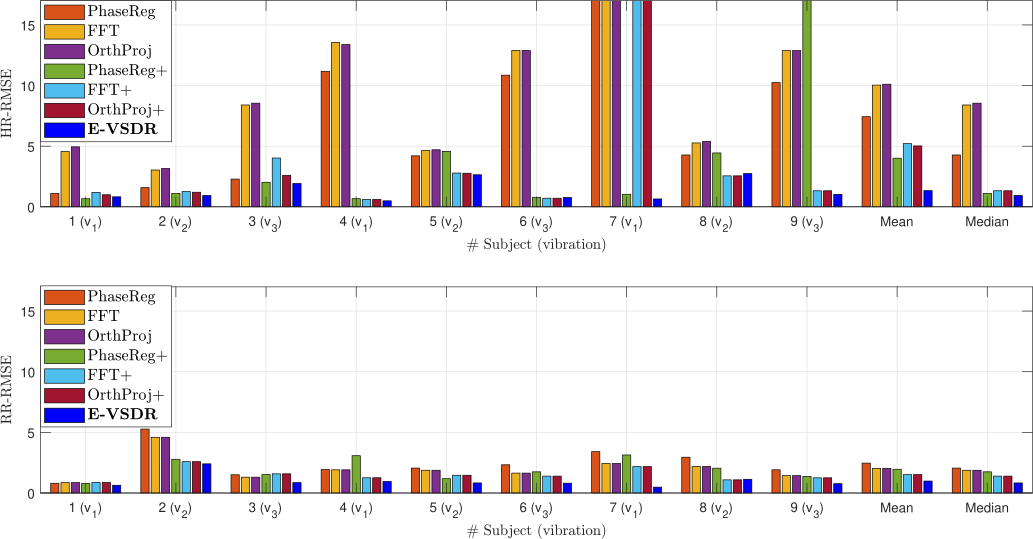}}
\end{center} 
\vspace{-0.2cm}
\caption{NCVSM performance plots for multi-person human trials c$4$. \textbf{(a)} Average empirical CDFs for HR and RR estimations. \textbf{(b)} RMSE scores: subject-wise with class average and median. For both HR and RR estimations, the proposed E-VSDR outperformed the compared methods. In terms of AeCDF, it surpassed for every error above $0.5$ [bpm], and in terms of RMSE, it achieved the lowest average and median values.}   
\label{fig:performance_t}
\vspace{-0.5cm}
\end{figure*}

Fig. \ref{fig:performance_AeCDF_t} shows the $9$ subjects-AeCDF of class c$4$. As in the phantom case, the E-VSDR consistently outperformed all other methods for thresholds above $0.5$ [bpm], achieving ASR2, ASR3 and ASR4 accuracies of $87.10\%$, $94.12\%$ and $95.54\%$, respectively, for HR estimation and $94.14\%$, $98.12\%$ and $98.69\%$, respectively, for RR estimation. Detailed values for the compared methods are provided in Table \ref{table:accuracy}. One can notice a considerable difference in performance compared to the other techniques, especially in the more challenging task of HR monitoring due to the weak heartbeat signature, in favor of the proposed approach. Another noteworthy observation is that in contrast to the phantom trials c$3$, the refinements applied to the competing methods (\textit{PhaseReg+}, \textit{FFT+} and \textit{OrthProj+}) yielded performance improvements for both HR and RR estimations. However, these enhancements remained insufficient to surpass the performance of the proposed approach. 

Finally, Fig. \ref{fig:performance_RMSE_t} illustrates the HR-RMSE and RR-RMSE distributions for each NCVSM method in class c$4$. Similar to the phantom trials c$3$, the proposed E-VSDR achieved the lowest average and median RMSE values for both HR and RR estimations. Specifically, the class ARMSE was as low as $1.33$ and $0.98$ for HR and RR estimations, respectively, with the values for the compared methods presented in Table \ref{table:accuracy}. Furthermore, the refinements applied in \textit{PhaseReg+}, \textit{FFT+}, and \textit{OrthProj+} resulted in improved performance for both HR and RR estimations, as observed for both the average and median metrics. However, even with these enhancements, these methods did not outperform the RMSE scores of the E-VSDR.

The obtained results underline the robustness of the E-VSDR method, which uniquely integrates a dictionary-based recovery with prior knowledge of cardiopulmonary activity to accurately estimate heartbeat and respiratory rates, even in the presence of considerable noise and interfering harmonics, in various experimental setups.

\begin{figure*}[htbp!]  
\begin{center}
\subfigure{\includegraphics[width=0.9\textwidth]{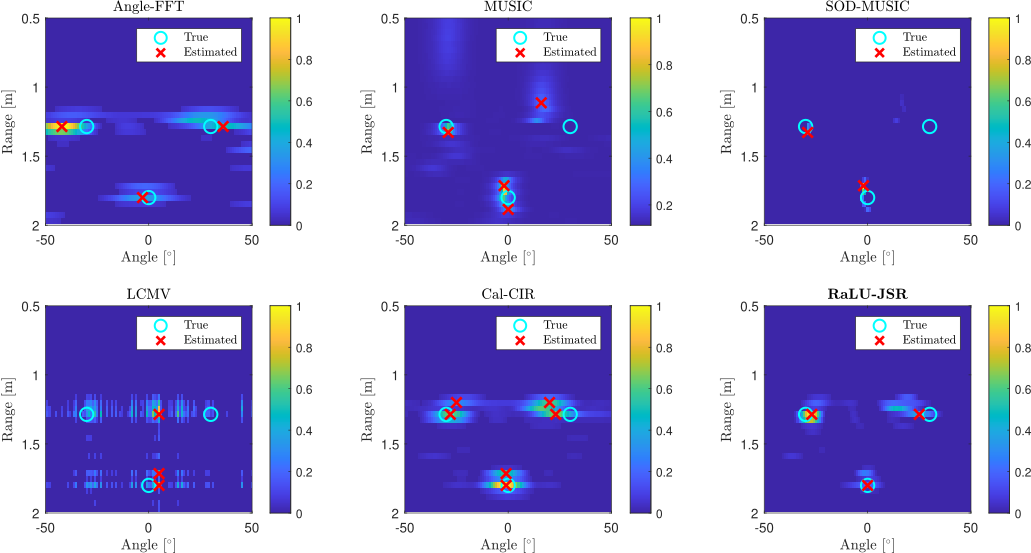}}\vspace{-0.2cm}
\end{center} 
\vspace{-0.2cm}
\caption{Localization maps of multi-person human trial - class c$4$, trial $\#3$. The maps were produced by \textit{Angle-FFT} \cite{ahmad2018vital,gao2019experiments,han2021detection}, \textit{MUSIC} \cite{konno2014experimental,kim2020low}, \textit{SOD-MUSIC} \cite{xiong2022vital}, \textit{LCMV} \cite{wang2021multi}, \textit{cal-CIR} \cite{wang2021driver} and the proposed RaLU-JSR (Algorithm \ref{alg_loc_3D_JSR_FISTA}). The X and O signs denote the estimated and true locations of humans, respectively. Only the proposed RaLU-JSR approach properly detects and positions all $3$ subjects in the specified scenario. }   
\label{fig:MP_loc_t}
\vspace{-0.4cm}
\end{figure*}

\section{Conclusion}
\label{sec:Conclusion}
This study introduces a comprehensive framework for multi-person localization and NCVSM using MIMO ULA FMCW radar, addressing critical challenges in cluttered, real-world environments. The contributions are twofold. First, we designed a custom hardware phantom capable of accurately replicating the cardiopulmonary dynamics of multiple individuals. This phantom provides a realistic and repeatable testbed for validating radar systems and algorithms, bridging the gap between theoretical development and practical deployment in clinical and home-care settings. Second, leveraging insights from the phantom validation process, we developed two novel algorithms: the RaLU-JSR method for multi-person localization, which exploits joint sparsity and cardiopulmonary properties, and the E-VSDR approach for continuous, harmonics-resilient vital signs estimation, effectively mitigating interference from respiration harmonics through a tailored dictionary-based method. The proposed framework was rigorously evaluated through both phantom trials and human trials, with results demonstrating the potential of the framework to advance NCVSM by providing robust, and accurate monitoring solutions, even in complex, cluttered environments. These findings offer a transformative step toward practical radar-based healthcare monitoring.  

\bibliographystyle{IEEEtran}
\bibliography{Main}

\begin{thebibliography}{10}
\providecommand{\url}[1]{#1}
\csname url@samestyle\endcsname
\providecommand{\newblock}{\relax}
\providecommand{\bibinfo}[2]{#2}
\providecommand{\BIBentrySTDinterwordspacing}{\spaceskip=0pt\relax}
\providecommand{\BIBentryALTinterwordstretchfactor}{4}
\providecommand{\BIBentryALTinterwordspacing}{\spaceskip=\fontdimen2\font plus
\BIBentryALTinterwordstretchfactor\fontdimen3\font minus \fontdimen4\font\relax}
\providecommand{\BIBforeignlanguage}[2]{{%
\expandafter\ifx\csname l@#1\endcsname\relax
\typeout{** WARNING: IEEEtran.bst: No hyphenation pattern has been}%
\typeout{** loaded for the language `#1'. Using the pattern for}%
\typeout{** the default language instead.}%
\else
\language=\csname l@#1\endcsname
\fi
#2}}
\providecommand{\BIBdecl}{\relax}
\BIBdecl

\bibitem{brekke2019value}
I.~J. Brekke, L.~H. Puntervoll, P.~B. Pedersen, J.~Kellett, and M.~Brabrand, ``The value of vital sign trends in predicting and monitoring clinical deterioration: A systematic review,'' \emph{PloS one}, vol.~14, no.~1, p. e0210875, 2019.

\bibitem{thakor1985electrode}
N.~Thakor and J.~Webster, ``Electrode studies for the long-term ambulatory ecg,'' \emph{Medical and Biological Engineering and Computing}, vol.~23, pp. 116--121, 1985.

\bibitem{ni2021automated}
X.~Ni, W.~Ouyang, H.~Jeong, J.-T. Kim, A.~Tzavelis, A.~Mirzazadeh, C.~Wu, J.~Y. Lee, M.~Keller, C.~K. Mummidisetty \emph{et~al.}, ``Automated, multiparametric monitoring of respiratory biomarkers and vital signs in clinical and home settings for covid-19 patients,'' \emph{Proceedings of the National Academy of Sciences}, vol. 118, no.~19, p. e2026610118, 2021.

\bibitem{brownsell1999future}
S.~Brownsell, G.~Williams, D.~A. Bradley, R.~Bragg, P.~Catlin, and J.~Carlier, ``Future systems for remote health care,'' \emph{Journal of Telemedicine and Telecare}, vol.~5, no.~3, pp. 141--152, 1999.

\bibitem{boric2002wireless}
O.~Boric-Lubeke and V.~M. Lubecke, ``Wireless house calls: using communications technology for health care and monitoring,'' \emph{IEEE Microwave Magazine}, vol.~3, no.~3, pp. 43--48, 2002.

\bibitem{nangalia2010health}
V.~Nangalia, D.~R. Prytherch, and G.~B. Smith, ``Health technology assessment review: Remote monitoring of vital signs-current status and future challenges,'' \emph{Critical Care}, vol.~14, pp. 1--8, 2010.

\bibitem{ceballos2015psychosocial}
P.~Ceballos-V{\'a}squez, G.~Rolo-Gonz{\'a}lez, E.~H{\'e}rnandez-Fernaud, D.~D{\'\i}az-Cabrera, T.~Paravic-Klijn, and M.~Burgos-Moreno, ``Psychosocial factors and mental work load: a reality perceived by nurses in intensive care units,'' \emph{Revista latino-americana de enfermagem}, vol.~23, no.~2, pp. 315--322, 2015.

\bibitem{fioranelli2019radar}
F.~Fioranelli, J.~Le~Kernec, and S.~A. Shah, ``Radar for health care: Recognizing human activities and monitoring vital signs,'' \emph{IEEE Potentials}, vol.~38, no.~4, pp. 16--23, 2019.

\bibitem{swaroop2019health}
K.~N. Swaroop, K.~Chandu, R.~Gorrepotu, and S.~Deb, ``A health monitoring system for vital signs using iot,'' \emph{Internet of Things}, vol.~5, pp. 116--129, 2019.

\bibitem{xiao2006ka}
Y.~Xiao, J.~Lin, O.~Boric-Lubecke, and V.~M. Lubecke, ``A ka-band low power doppler radar system for remote detection of cardiopulmonary motion,'' in \emph{2005 IEEE Engineering in Medicine and Biology 27th Annual Conference}.\hskip 1em plus 0.5em minus 0.4em\relax IEEE, 2006, pp. 7151--7154.

\bibitem{gu2012accurate}
C.~Gu, R.~Li, H.~Zhang, A.~Y. Fung, C.~Torres, S.~B. Jiang, and C.~Li, ``Accurate respiration measurement using dc-coupled continuous-wave radar sensor for motion-adaptive cancer radiotherapy,'' \emph{IEEE Transactions on biomedical engineering}, vol.~59, no.~11, pp. 3117--3123, 2012.

\bibitem{zhao2018noncontact}
H.~Zhao, H.~Hong, D.~Miao, Y.~Li, H.~Zhang, Y.~Zhang, C.~Li, and X.~Zhu, ``A noncontact breathing disorder recognition system using 2.4-ghz digital-if doppler radar,'' \emph{IEEE journal of biomedical and health informatics}, vol.~23, no.~1, pp. 208--217, 2018.

\bibitem{alizadeh2019remote}
M.~Alizadeh, G.~Shaker, J.~C.~M. De~Almeida, P.~P. Morita, and S.~Safavi-Naeini, ``Remote monitoring of human vital signs using mm-wave fmcw radar,'' \emph{IEEE Access}, vol.~7, pp. 54\,958--54\,968, 2019.

\bibitem{sacco2020fmcw}
G.~Sacco, E.~Piuzzi, E.~Pittella, and S.~Pisa, ``An fmcw radar for localization and vital signs measurement for different chest orientations,'' \emph{Sensors}, vol.~20, no.~12, p. 3489, 2020.

\bibitem{antolinos2020cardiopulmonary}
E.~Antolinos, F.~Garc{\'\i}a-Rial, C.~Hern{\'a}ndez, D.~Montesano, J.~I. Godino-Llorente, and J.~Grajal, ``Cardiopulmonary activity monitoring using millimeter wave radars,'' \emph{Remote Sensing}, vol.~12, no.~14, p. 2265, 2020.

\bibitem{kim2018low}
S.~Kim and K.-K. Lee, ``Low-complexity joint extrapolation-music-based 2-d parameter estimator for vital fmcw radar,'' \emph{IEEE sensors journal}, vol.~19, no.~6, pp. 2205--2216, 2018.

\bibitem{turppa2020vital}
E.~Turppa, J.~M. Kortelainen, O.~Antropov, and T.~Kiuru, ``Vital sign monitoring using fmcw radar in various sleeping scenarios,'' \emph{Sensors}, vol.~20, no.~22, p. 6505, 2020.

\bibitem{eder2023sparsity}
Y.~Eder and Y.~C. Eldar, ``Sparsity-based multi-person non-contact vital signs monitoring via fmcw radar,'' \emph{IEEE journal of biomedical and health informatics}, vol.~27, no.~6, pp. 2806--2817, 2023.

\bibitem{eder2023sparse}
Y.~Eder, Z.~Liu, and Y.~C. Eldar, ``Sparse non-contact multiple people localization and vital signs monitoring via fmcw radar,'' in \emph{ICASSP 2023-2023 IEEE International Conference on Acoustics, Speech and Signal Processing (ICASSP)}.\hskip 1em plus 0.5em minus 0.4em\relax IEEE, 2023, pp. 1--5.

\bibitem{li2021through}
Z.~Li, T.~Jin, Y.~Dai, and Y.~Song, ``Through-wall multi-subject localization and vital signs monitoring using uwb mimo imaging radar,'' \emph{Remote Sensing}, vol.~13, no.~15, p. 2905, 2021.

\bibitem{feng2021multitarget}
C.~Feng, X.~Jiang, M.-G. Jeong, H.~Hong, C.-H. Fu, X.~Yang, E.~Wang, X.~Zhu, and X.~Liu, ``Multitarget vital signs measurement with chest motion imaging based on mimo radar,'' \emph{IEEE Transactions on Microwave Theory and Techniques}, vol.~69, no.~11, pp. 4735--4747, 2021.

\bibitem{juan2022distributed}
P.-H. Juan, C.-Y. Chueh, and F.-K. Wang, ``Distributed mimo cw radar for locating multiple people and detecting their vital signs,'' \emph{IEEE Transactions on Microwave Theory and Techniques}, vol.~71, no.~3, pp. 1312--1325, 2022.

\bibitem{hsieh2024multiperson}
C.-H. Hsieh and P.-H. Tseng, ``Multiperson localization and vital signs estimation using mmwave mimo radar,'' \emph{IEEE Transactions on Microwave Theory and Techniques}, 2024.

\bibitem{li2021signal}
X.~Li, X.~Wang, Q.~Yang, and S.~Fu, ``Signal processing for tdm mimo fmcw millimeter-wave radar sensors,'' \emph{IEEE Access}, vol.~9, pp. 167\,959--167\,971, 2021.

\bibitem{xu2022simultaneous}
Z.~Xu, C.~Shi, T.~Zhang, S.~Li, Y.~Yuan, C.-T.~M. Wu, Y.~Chen, and A.~Petropulu, ``Simultaneous monitoring of multiple people’s vital sign leveraging a single phased-mimo radar,'' \emph{IEEE Journal of Electromagnetics, RF and Microwaves in Medicine and Biology}, vol.~6, no.~3, pp. 311--320, 2022.

\bibitem{mercuri2021enabling}
M.~Mercuri, Y.~Lu, S.~Polito, F.~Wieringa, Y.-H. Liu, A.-J. van~der Veen, C.~Van~Hoof, and T.~Torfs, ``Enabling robust radar-based localization and vital signs monitoring in multipath propagation environments,'' \emph{IEEE Transactions on Biomedical Engineering}, vol.~68, no.~11, pp. 3228--3240, 2021.

\bibitem{paterniani2023radar}
G.~Paterniani, D.~Sgreccia, A.~Davoli, G.~Guerzoni, P.~Di~Viesti, A.~C. Valenti, M.~Vitolo, G.~M. Vitetta, and G.~Boriani, ``Radar-based monitoring of vital signs: A tutorial overview,'' \emph{Proceedings of the IEEE}, vol. 111, no.~3, pp. 277--317, 2023.

\bibitem{mercuri2019vital}
M.~Mercuri, I.~R. Lorato, Y.-H. Liu, F.~Wieringa, C.~V. Hoof, and T.~Torfs, ``Vital-sign monitoring and spatial tracking of multiple people using a contactless radar-based sensor,'' \emph{Nature Electronics}, vol.~2, no.~6, pp. 252--262, 2019.

\bibitem{fatadin2008compensation}
I.~Fatadin, S.~J. Savory, and D.~Ives, ``Compensation of quadrature imbalance in an optical qpsk coherent receiver,'' \emph{IEEE Photonics Technology Letters}, vol.~20, no.~20, pp. 1733--1735, 2008.

\bibitem{mahendra2020compensation}
R.~Mahendra, S.~K. Mohammed, and R.~K. Mallik, ``Compensation of receiver iq imbalance in mm-wave hybrid beamforming systems,'' in \emph{2020 IEEE 92nd Vehicular Technology Conference (VTC2020-Fall)}.\hskip 1em plus 0.5em minus 0.4em\relax IEEE, 2020, pp. 1--6.

\bibitem{ahmad2018vital}
A.~Ahmad, J.~C. Roh, D.~Wang, and A.~Dubey, ``Vital signs monitoring of multiple people using a fmcw millimeter-wave sensor,'' in \emph{2018 IEEE Radar Conference (RadarConf18)}.\hskip 1em plus 0.5em minus 0.4em\relax IEEE, 2018, pp. 1450--1455.

\bibitem{gao2019experiments}
X.~Gao, G.~Xing, S.~Roy, and H.~Liu, ``Experiments with mmwave automotive radar test-bed,'' in \emph{2019 53rd Asilomar conference on signals, systems, and computers}.\hskip 1em plus 0.5em minus 0.4em\relax IEEE, 2019, pp. 1--6.

\bibitem{han2021detection}
K.~Han and S.~Hong, ``Detection and localization of multiple humans based on curve length of i/q signal trajectory using mimo fmcw radar,'' \emph{IEEE Microwave and Wireless Components Letters}, vol.~31, no.~4, pp. 413--416, 2021.

\bibitem{konno2014experimental}
K.~Konno, M.~Nango, N.~Honma, K.~Nishimori, N.~Takemura, and T.~Mitsui, ``Experimental evaluation of estimating living-body direction using array antenna for multipath environment,'' \emph{IEEE Antennas and Wireless Propagation Letters}, vol.~13, pp. 718--721, 2014.

\bibitem{kim2020low}
B.-s. Kim, Y.~Jin, J.~Lee, and S.~Kim, ``Low-complexity music-based direction-of-arrival detection algorithm for frequency-modulated continuous-wave vital radar,'' \emph{Sensors}, vol.~20, no.~15, p. 4295, 2020.

\bibitem{xiong2022vital}
J.~Xiong, H.~Hong, L.~Xiao, E.~Wang, and X.~Zhu, ``Vital signs detection with difference beamforming and orthogonal projection filter based on simo-fmcw radar,'' \emph{IEEE Transactions on Microwave Theory and Techniques}, vol.~71, no.~1, pp. 83--92, 2022.

\bibitem{wang2021multi}
Y.~Wang, Y.~Shui, X.~Yang, Z.~Li, and W.~Wang, ``Multi-target vital signs detection using frequency-modulated continuous wave radar,'' \emph{EURASIP Journal on Advances in Signal Processing}, vol. 2021, pp. 1--19, 2021.

\bibitem{wang2021driver}
F.~Wang, X.~Zeng, C.~Wu, B.~Wang, and K.~R. Liu, ``Driver vital signs monitoring using millimeter wave radio,'' \emph{IEEE Internet of Things Journal}, vol.~9, no.~13, pp. 11\,283--11\,298, 2021.

\bibitem{park2007arctangent}
B.-K. Park, O.~Boric-Lubecke, and V.~M. Lubecke, ``Arctangent demodulation with dc offset compensation in quadrature doppler radar receiver systems,'' \emph{IEEE transactions on Microwave theory and techniques}, vol.~55, no.~5, pp. 1073--1079, 2007.

\bibitem{adib2015smart}
F.~Adib, H.~Mao, Z.~Kabelac, D.~Katabi, and R.~C. Miller, ``Smart homes that monitor breathing and heart rate,'' in \emph{Proceedings of the 33rd annual ACM conference on human factors in computing systems}, 2015, pp. 837--846.

\bibitem{bakhtiari2011compact}
S.~Bakhtiari, T.~W. Elmer, N.~M. Cox, N.~Gopalsami, A.~C. Raptis, S.~Liao, I.~Mikhelson, and A.~V. Sahakian, ``Compact millimeter-wave sensor for remote monitoring of vital signs,'' \emph{IEEE Transactions on Instrumentation and Measurement}, vol.~61, no.~3, pp. 830--841, 2011.

\bibitem{islam2019programmable}
S.~M.~M. Islam, B.~Tomota, A.~Sylvester, and V.~M. Lubecke, ``A programmable robotic phantom to simulate the dynamic respiratory motions of humans for continuous identity authentication,'' in \emph{2019 IEEE Asia-Pacific Microwave Conference (APMC)}.\hskip 1em plus 0.5em minus 0.4em\relax IEEE, 2019, pp. 1408--1410.

\bibitem{marty2023investigation}
S.~Marty, F.~Pantanella, A.~Ronco, K.~Dheman, and M.~Magno, ``Investigation of mmwave radar technology for non-contact vital sign monitoring,'' in \emph{2023 IEEE International Symposium on Medical Measurements and Applications (MeMeA)}.\hskip 1em plus 0.5em minus 0.4em\relax IEEE, 2023, pp. 1--6.

\bibitem{schellenberger2020dataset}
S.~Schellenberger, K.~Shi, T.~Steigleder, A.~Malessa, F.~Michler, L.~Hameyer, N.~Neumann, F.~Lurz, R.~Weigel, C.~Ostgathe \emph{et~al.}, ``A dataset of clinically recorded radar vital signs with synchronised reference sensor signals,'' \emph{Scientific data}, vol.~7, no.~1, p. 291, 2020.

\bibitem{iovescu2017fundamentals}
C.~Iovescu and S.~Rao, ``The fundamentals of millimeter wave sensors,'' \emph{Texas Instruments}, pp. 1--8, 2017.

\bibitem{ruppert1994multivariate}
D.~Ruppert and M.~P. Wand, ``Multivariate locally weighted least squares regression,'' \emph{The annals of statistics}, pp. 1346--1370, 1994.

\bibitem{rossman2019rapid}
U.~Rossman, R.~Tenne, O.~Solomon, I.~Kaplan-Ashiri, T.~Dadosh, Y.~C. Eldar, and D.~Oron, ``Rapid quantum image scanning microscopy by joint sparse reconstruction,'' \emph{Optica}, vol.~6, no.~10, pp. 1290--1296, 2019.

\bibitem{beck2009fast}
A.~Beck and M.~Teboulle, ``A fast iterative shrinkage-thresholding algorithm for linear inverse problems,'' \emph{SIAM journal on imaging sciences}, vol.~2, no.~1, pp. 183--202, 2009.

\bibitem{palomar2010convex}
D.~P. Palomar and Y.~C. Eldar, \emph{Convex optimization in signal processing and communications}.\hskip 1em plus 0.5em minus 0.4em\relax Cambridge university press, 2010.

\bibitem{jalil2016analysis}
A.~Jalil, H.~Yousaf, and M.~I. Baig, ``Analysis of cfar techniques,'' in \emph{2016 13th International Bhurban Conference on Applied Sciences and Technology (IBCAST)}.\hskip 1em plus 0.5em minus 0.4em\relax IEEE, 2016, pp. 654--659.

\bibitem{ernst1999impedance}
J.~M. Ernst, D.~A. Litvack, D.~L. Lozano, J.~T. Cacioppo, and G.~G. Berntson, ``Impedance pneumography: Noise as signal in impedance cardiography,'' \emph{Psychophysiology}, vol.~36, no.~3, pp. 333--338, 1999.

\bibitem{sherwood1990methodological}
A.~Sherwood, M.~T. Allen, J.~Fahrenberg, R.~M. Kelsey, W.~R. Lovallo, and L.~J. Van~Doornen, ``Methodological guidelines for impedance cardiography,'' \emph{Psychophysiology}, vol.~27, no.~1, pp. 1--23, 1990.

\bibitem{naughton1998pathophysiology}
M.~Naughton, ``Pathophysiology and treatment of cheyne-stokes respiration,'' \emph{Thorax}, vol.~53, no.~6, pp. 514--518, 1998.

\bibitem{arduino_nano}
\BIBentryALTinterwordspacing
{Arduino Documentation}, ``{Arduino Nano Hardware Documentation},'' 2024. [Online]. Available: \url{https://docs.arduino.cc/hardware/nano/}
\BIBentrySTDinterwordspacing

\bibitem{vibration_generator_3bscientific}
\BIBentryALTinterwordspacing
{3B Scientific}, ``{Vibration Generator: For Investigating Oscillations and Resonances},'' 2024. [Online]. Available: \url{https://www.3bscientific.com/il/vibration-generator-for-investigating-\break oscillations-and-resonances-1000701-u56001-3b-scientific,p_576_1977.html}
\BIBentrySTDinterwordspacing

\bibitem{chladni_plate_3bscientific}
\BIBentryALTinterwordspacing
{3B Scientific (2)}, ``{Chladni Plate, Circular: For Generating Acoustically Excited Chladni Figures},'' 2024. [Online]. Available: \url{https://www.3bscientific.com/il/chladni-plate-circular-for-generating-\break acoustically-excited-chladni-figures-1000705-u56005-3b-scientific,p_576_1981.html}
\BIBentrySTDinterwordspacing

\bibitem{gtec_biosignal_amplifier}
\BIBentryALTinterwordspacing
{g.tec medical engineering GmbH}, ``{g.HIamp - 256-Channel Biosignal Amplifier},'' 2024. [Online]. Available: \url{https://www.gtec.at/product/g-hiamp-256-channel-biosignal-amplifier/}
\BIBentrySTDinterwordspacing

\bibitem{gtec_body_sensors}
\BIBentryALTinterwordspacing
{g.tec medical engineering GmbH (2)}, ``{Body Sensors: Wireless and Wearable Sensors for Physiological Signals},'' 2024. [Online]. Available: \url{https://www.gtec.at/product/body-sensors/}
\BIBentrySTDinterwordspacing

\bibitem{ti_iwr1443boost}
\BIBentryALTinterwordspacing
{Texas Instruments}, ``{IWR1443BOOST: mmWave Sensor Evaluation Module},'' 2024. [Online]. Available: \url{https://www.ti.com/tool/IWR1443BOOST}
\BIBentrySTDinterwordspacing

\bibitem{ti_dca1000evm}
\BIBentryALTinterwordspacing
{Texas Instruments (2)}, ``{DCA1000EVM: mmWave Data Capture Card Evaluation Module},'' 2024. [Online]. Available: \url{https://www.ti.com/tool/DCA1000EVM}
\BIBentrySTDinterwordspacing

\end{thebibliography}

\vspace{-1cm}
\begin{IEEEbiography}[{\includegraphics[width=1in,height=1.25in,clip,keepaspectratio]{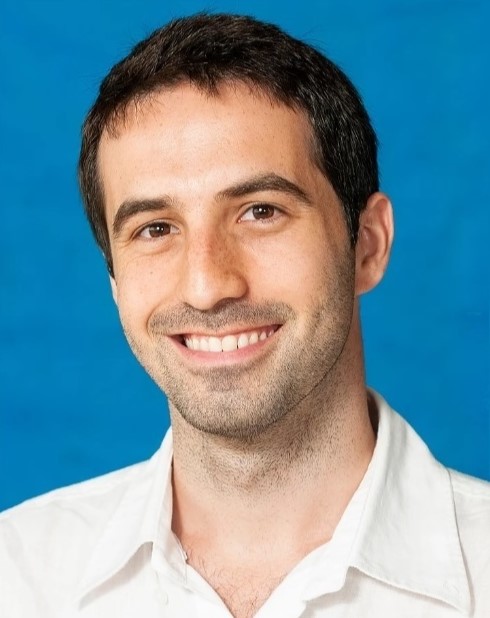}}]{Yonathan Eder} (Graduate Student Member, IEEE) received the B.Sc. and M.Sc. degrees in electrical and computer engineering from the Ben-Gurion University of the Negev, Beer-Sheva, Israel, in 2018 and 2021, respectively. He is currently pursuing the Ph.D. degree at the Faculty of Mathematics and Computer Science, Weizmann Institute of Science, Rehovot, Israel.  
Mr. Eder is a co-inventor on multiple patents (pending approval) related to radar technology for biomedical applications. His research interests include signal processing, model-based deep learning, and non-contact physiological sensing using radar systems. His work has contributed to advancing radar-based solutions for non-contact health monitoring, bridging theoretical innovation with practical applications.  
\end{IEEEbiography}

\begin{IEEEbiography}[{\includegraphics[width=1in,height=1.25in,clip,keepaspectratio]{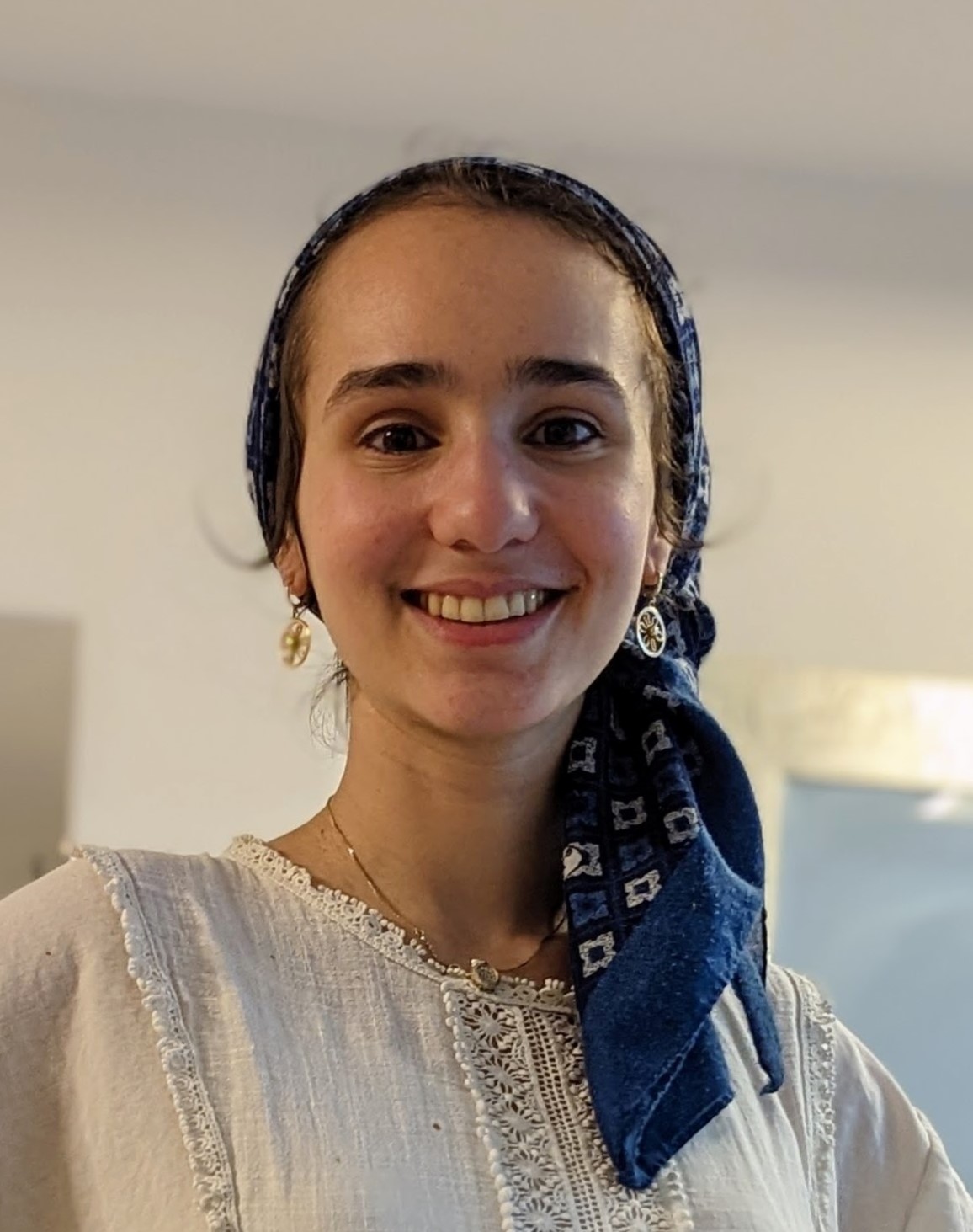}}]{Emma Zagoury} received the B.Sc. degree in biomedical engineering from Tel Aviv University, Tel Aviv, Israel, where she is currently pursuing the M.Sc. degree in electrical engineering. She worked as an Algorithm Engineer at SAMPL Lab, headed by Prof. Yonina C. Eldar, at the Faculty of Mathematics and Computer Science, Weizmann Institute of Science, Rehovot, Israel. Her research interests include signal processing, radar technologies, and the integration of biomedical and electrical engineering for health monitoring.
\end{IEEEbiography}

\begin{IEEEbiography}[{\includegraphics[width=1in,height=1.25in,clip,keepaspectratio]{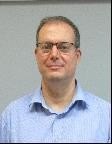}}]{Shlomi Savariego} received the B.Sc degree in electrical and electronics engineering and an MBA in technology and information systems management from Tel Aviv University, Tel Aviv, Israel, in 1993 and 2009, respectively. He is currently a Hardware Engineer at SAMPL Lab, headed by Prof. Yonina C. Eldar at the Faculty of Mathematics and Computer Science, Weizmann Institute of Science, Rehovot, Israel. His work focuses on the development of advanced systems in ultrasound, radar, and demonstrative platforms. With over 30 years of experience, he has specialized in real-time systems for cellular communication, control, and sensing.
\end{IEEEbiography}

\begin{IEEEbiography}[{\includegraphics[width=1in,height=1.25in,clip,keepaspectratio]{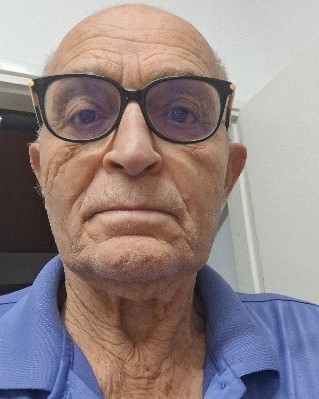}}]{Moshe Namer} received the B.Sc. degree in electrical engineering from the Technion - Israel Institute of Technology, Haifa, Israel, in 1984. He has extensive experience in managing and leading advanced engineering projects. As the head of the Technion Communication Laboratory, he oversaw the implementation of projects spanning a wide range of frequencies and developed expertise in analog hardware applications.

\end{IEEEbiography}

\begin{IEEEbiography}[{\includegraphics[width=1in,height=1.25in,clip,keepaspectratio]{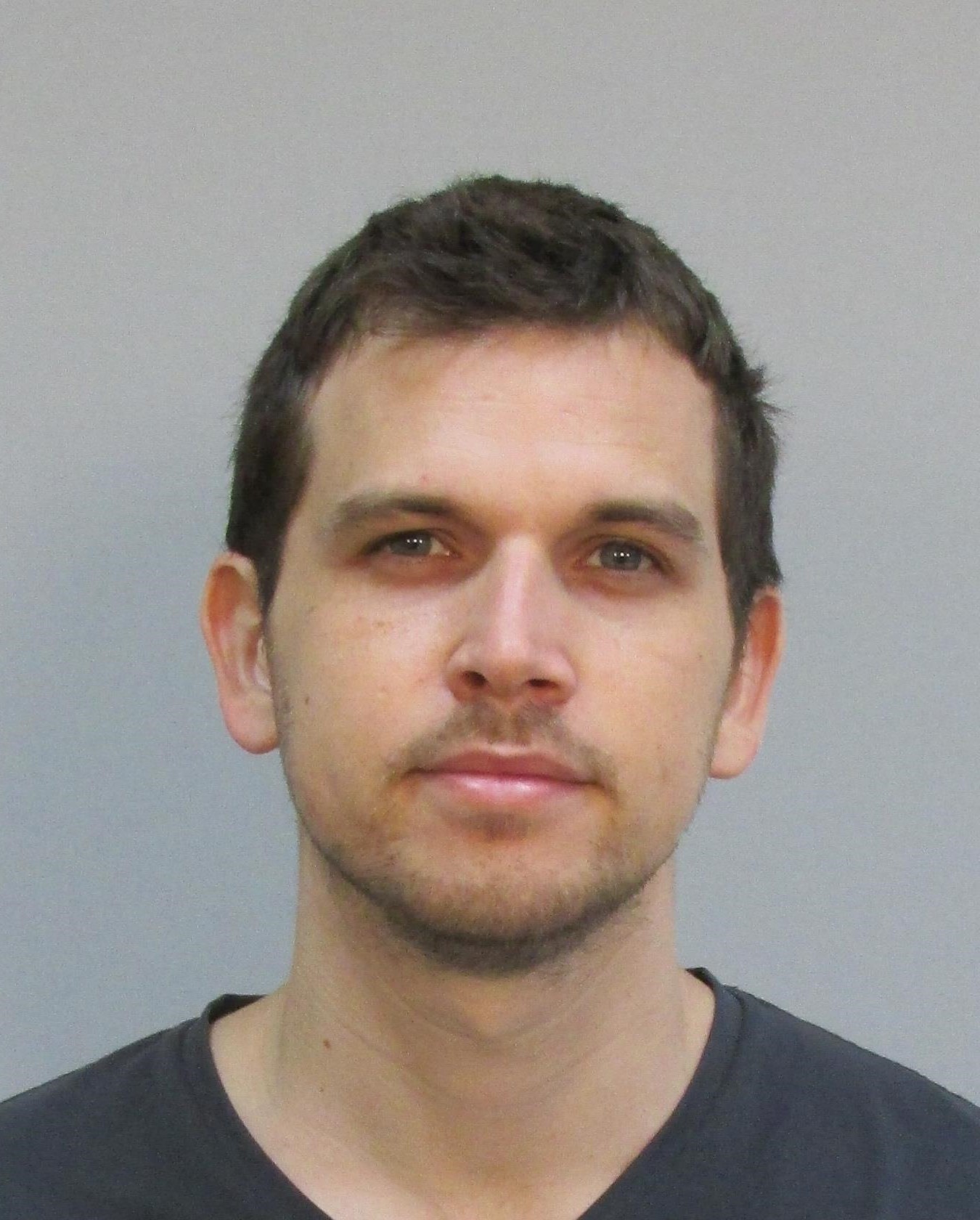}}]{Oded Cohen} received the B.Sc. degree in electrical and computer engineering from the Ben Gurion University of the Negev, Beer-Sheva, Israel, in 2019. He works as an Algorithm Engineer at SAMPL Lab, headed by Prof. Yonina C. Eldar at the Faculty of Mathematics and Computer Science, Weizmann Institute of Science, Rehovot, Israel.
\end{IEEEbiography}

\begin{IEEEbiography}[{\includegraphics[width=1in,height=1.25in,clip,keepaspectratio]{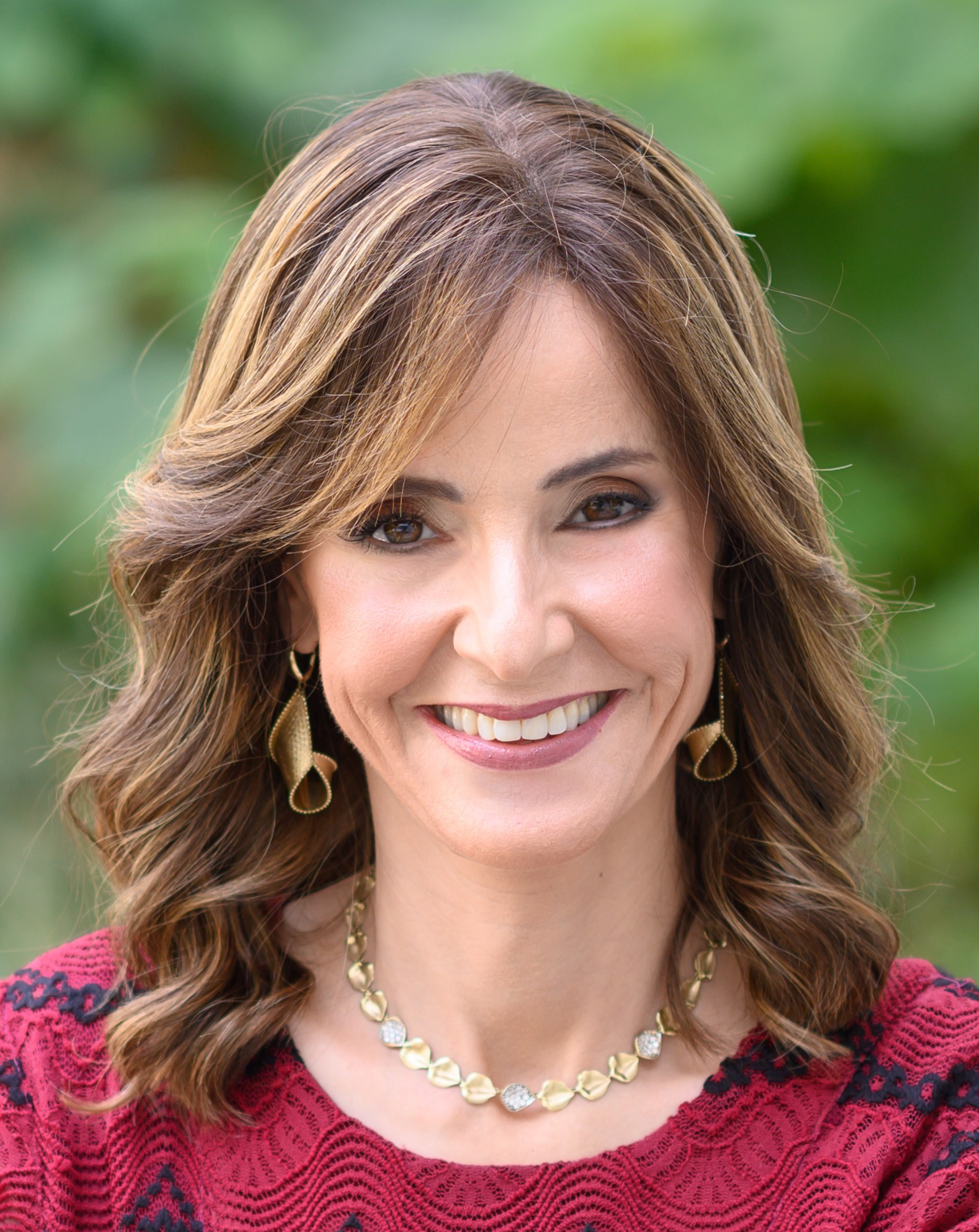}}]{Yonina C. Eldar} (Fellow, IEEE) is a Professor in the Department of Mathematics and Computer Science, Weizmann Institute of Science, Rehovot, Israel where she heads the center for Biomedical Engineering and Signal Processing and holds the Dorothy and Patrick Gorman Professorial Chair. She is also a Visiting Professor at MIT, a Visiting Scientist at the Broad Institute, and an Adjunct Professor at Duke University and was a Visiting Professor at Stanford.  She is a member of the Israel Academy of Sciences and Humanities, an IEEE Fellow and a EURASIP Fellow. She received the B.Sc. degree in physics and the B.Sc. degree in electrical engineering from Tel-Aviv University, and the Ph.D. degree in electrical engineering and computer science from MIT, in 2002. She has received many awards for excellence in research and teaching, including the IEEE Signal Processing Society Technical Achievement Award (2013), the IEEE/AESS Fred Nathanson Memorial Radar Award (2014) and the IEEE Kiyo Tomiyasu Award (2016). She was a Horev Fellow of the Leaders in Science and Technology program at the Technion and an Alon Fellow. She received the Michael Bruno Memorial Award from the Rothschild Foundation, the Weizmann Prize for Exact Sciences, the Wolf Foundation Krill Prize for Excellence in Scientific Research, the Henry Taub Prize for Excellence in Research (twice), the Hershel Rich Innovation Award (three times), and the Award for Women with Distinguished Contributions. She received several best paper awards and best demo awards together with her research students and colleagues, was selected as one of the 50 most influential women in Israel, and was a member of the Israel Committee for Higher Education. She is the Editor in Chief of Foundations and Trends in Signal Processing, a member of several IEEE Technical Committees and Award Committees, and heads the Committee for Promoting Gender Fairness in Higher Education Institutions in Israel.
\end{IEEEbiography}

\end{document}